\documentclass[twocolumn,groupedaddress,aps,pra,10pt,floatfix]{revtex4-2}
\usepackage[colorlinks,linkcolor=blue,anchorcolor=blue,citecolor=blue,bookmarksnumbered]{hyperref}
\usepackage{graphicx,epsfig,float,epstopdf}
\usepackage{amssymb,amsmath,subeqnarray,mathrsfs,amsfonts,bm,makecell,cases,color,extpfeil,lipsum}
\usepackage[normalem]{ulem}
\usepackage{orcidlink,soul,lineno}
\usepackage{graphicx}
\usepackage{dcolumn}
\usepackage{bm}
 
\allowdisplaybreaks

\begin{document}

\title{Stable (2+1)-dimensional soliton and breather molecules in a
cold Rydberg atomic gas}

\author{Lu Qin$^{1}$\orcidlink{0000-0001-6977-2680}}
\author{Hairu Zhai$^{1}$}
\author{Lu Liu$^{1}$}
\author{Yingying Zhang$^{1}$\textsuperscript{*}}
\author{Zeyun Shi$^{2}$\textsuperscript{$\dagger$}\orcidlink{0000-0003-4266-0691}}
\author{Zunlue Zhu$^{1}$}
\author{Xingdong Zhao$^{1}$\textsuperscript{$\ddagger$}%
\orcidlink{0000-0001-9533-7279}}
\author{Wu-Ming Liu$^{1,3}$}
\author{Boris A. Malomed$^{4,5}$}

\affiliation{$^{1}$School of Physics, Henan Normal University, Xinxiang 453007, China\\
$^2$School of Artificial Intelligence, Hubei University of Automotive Technology, Shiyan 442000, China\\
$^3$Beijing National Laboratory for Condensed Matter Physics, Institute of Physics, Chinese Academy of Sciences, Beijing 100190, China\\
$^4$
Department of Physical Electronics
School of Electrical and Computer Engineering Faculty of Engineering
Tel Aviv University Ramat Aviv, Tel Aviv 6997801, Israel\\
$^5$Instituto de Alta Investigación
Universidad de Tarapacá
Casilla 7D, Arica1 000000, Chile}

\thanks{yyzwuli415@163.com\\
$^{\dagger}$shizy@huat.edu.cn\\
$^{\ddagger}$phyzhxd@gmail.com}
\date{\today }

\begin{abstract}
We investigate the formation of stable (2+1)-dimensional [(2+1)D] spatial-domain
optical soliton molecules (SMs) and breather molecules (BMs) in a gas of
Rydberg atoms, highlighting the role of the nonlocal nonlinearity,
which is generated by the electromagnetically induced transparency (EIT) in
the Rydberg medium. The setting supports diverse species of large-size
polygonal SMs, including rectangular and oblique rhombuses, checkerboard
cells, and hexagons. The analysis identifies two distinct formation regimes.
In the case of moderately nonlocality, the long-range interactions
alone stabilize the SMs in the static form. In contrast, in the strongly
nonlocal regime, initially imposed rotation is required to generate a
centrifugal force that counteracts the strong attraction, resulting in
stably rotating SMs. The rotation period can be controlled by adjusting the
system's parameters. Furthermore, appropriate initial velocities can induce
inherent breathing dynamics in the solitons, leading to the formation of
BMs. Tuning the initial velocity, one can control the evolution of SMs and
BMs and even realize their mutual conversion. Our study offers a new scheme
for engineering SMs and BMs, and suggests new possibilities for the design
of data processing and transmission in optical systems.
% {\color{blue}Our research has provided new solutions for designing soliton molecules with different configurations, made experimental predictions, and opened up new avenues for data processing and transmission in optical systems.}
\end{abstract}

\maketitle

% ---- 标题页 ----

\thispagestyle{titlepage}

% ---- 正文 ----

\section{Introduction}

% \textcolor[rgb]{1.00,0.00,0.50}{SM $\dashrightarrow$ SMs}

Solitons, alias self-localized nonlinear waves, have attracted great
attention in hydrodynamics and plasmas~\cite{KUZNETSOV1986103}, optics~\cite%
{Torruellas2001,Hasegawa1989,Akhmediev2005,malomedSolitonManagementPeriodic2006a,LEDERER20081,Dai2025_SC}%
, superconductivity~\cite{ustinovDynamicsSineGordonSolitons1992}, and
quantum systems~\cite%
{WOS:001096218600011,kivsharDynamicsSolitonsNearly1989,Sakkaravarthi2023}.
Nowadays, stable bound states~of solitons \cite%
{desyatnikovRotatingOpticalSoliton2002a,Malomed1991,Malomed1993,Skryabin,Tang-1,Komarov2009,Gelash2019} that arise from the balance of attractive and repulsive interactions
between adjacent ones, also known as soliton molecules (SMs)~\cite%
{Liu2023,guoUnveilingComplexitySpatiotemporal2023a,alkhawajaInteractionForcesTwodimensional2012,qin2022stable,peng2018build}%
, have become a new research frontier. Their unique properties offer a wide
range of potential applications in various physical systems, including
mode-locked fiber lasers~\cite{Zavyalov2009}, dispersion management~\cite%
{Boudje2013}, optical microresonators~\cite{weng_heteronuclear2020},
exciton-polariton superfluids~\cite{maitre2020}, Bose-Einstein condensates
(BEC)~\cite{Santos2012,Shirley2014}, etc. In particular, SMs provide
promising applications in optics, including the design of new laser schemes~%
\cite{tsatourian_polarisation_2013}, optical switches \cite%
{yinCoherentAtomicSoliton2011a,Kurtz2020,liu_demand_2022}, coherent
frequency combs~\cite{weng_heteronuclear2020}, advanced optical
telecommunications~\cite{mitschke_soliton_2016}, data encoding and
transmission \cite%
{rohrmannTwosolitonThreesolitonMolecules2013,Alamoudi2014,Liu2023,rohrmann2012solitons,marin2017microresonator,Dai2025,liu2025gain,yang2024phase}%
, etc.

Earlier work on SMs was confined to one-dimensional (1D) settings with local
nonlinearity, exemplified by temporal SMs in fiber lasers~\cite%
{liuDynamicEvolutionTemporal2010,wangFewlayerBismutheneFemtosecond2019,liVariousSolitonMolecules,qinObservationSolitonMolecules2018,Wang2019,Igbonacho2019,xu2019breather}%
. In local nonlinear systems, SMs form solely through short-range (contact)
interactions, making the creation of large-size SMs in 1D and the transition
to 2D and 3D a challenging problem, due to instabilities emerging in these
cases. Indeed, unlike 1D solitons, which normally appear as stable modes,
the stability of 2D and 3D solitons is a fundamental problem, as the usual
cubic self-focusing gives rise to the critical and supercritical collapse in
the 2D and 3D space, respectively \cite%
{chenOpticalSolitonsSelffocusing2018,askar2008effects,chiaoSelfTrappingOpticalBeams1964,lallemand1965self,garmire1966dynamics}%
. Therefore, devising physically viable stabilization mechanisms for 2D and
3D solitons and their bound states remains a major challenge. Nevertheless,
 research has confirmed the existence of stable multi-dimensional
solitons in relatively complex physical systems~\cite%
{snyderAccessibleSolitons1997,krolikowskiSolitonsNonlocalNonlinear2000,contiObservationOpticalSpatial2004}%
. Theoretically, 2D solitons have been demonstrated in media with competing
nonlinearities~\cite%
{kartashovRobustPropagationTwoColor2002,mihalacheRobustSolitonClusters2003},
as well as in systems involving local nonlinearity combined with an external
nonlinear potential~\cite{zengRobustDynamicsSoliton2023}. Experimentally,
stable 2D solitons have been observed in optical media~with 
cubic-quintic nonlinearity \cite%
{falcao-filhoRobustTwoDimensionalSpatial2013a}, and various types of
solitons-like modes (\textquotedblleft quantum droplets") have been
identified in BEC~\cite{cheineyBrightSolitonQuantum2018}. 
Multi-dimensional solitons and their bound states in the form of the
corresponding SMs\ remain a relevant topic for theoretical and experimental
studies \cite{malomed2022multidimensional}.

Recently, it has been shown that cold atomic gases steered by laser fields
provide an appropriate ground for studying multi-dimensional solitons~%
\cite%
{firstenbergNonlinearQuantumOptics2016,Gallagher2008,Saffman2010,bai_self-induced_2020,rotschild2006NP}%
. In particular, by coupling light to highly excited Rydberg states~\cite%
{firstenbergNonlinearQuantumOptics2016,murray2016quantum,pritchard2013nonlinear}%
, strong and long-range interactions between Rydberg atoms can be induced by
light fields through the effect of  electromagnetically induced
transparency (EIT)~\cite{Fleischhauer2005}, thus generating strong nonlocal
nonlinearity~\cite{Firstenberg2013,murray2016quantum}. Actually, the
Rydberg-Rydberg interactions are strong even at the single-photon level~\cite%
{pritchard2013nonlinear,firstenbergNonlinearQuantumOptics2016,murray2016quantum}%
. Photonic dimers and three-photon bound states have been observed under the
action of strong optical nonlinearities~\cite%
{Firstenberg2013,liang2018observation}. Such bound states of photons can be
viewed as quantum solitons, and they may be used for the implementation of
single-photon switching~\cite{baur2014single}, photon-photon gates~\cite%
{gorshkov2011photon}, and all-optical deterministic quantum logic~\cite%
{shahmoon2011strongly}, as well as for studies of many-body phenomena with
strongly correlated photons~\cite{chang2008crystallization}.

%===========================fig1===============================%
\begin{figure}[t]
\centering
\includegraphics[width=1\linewidth]{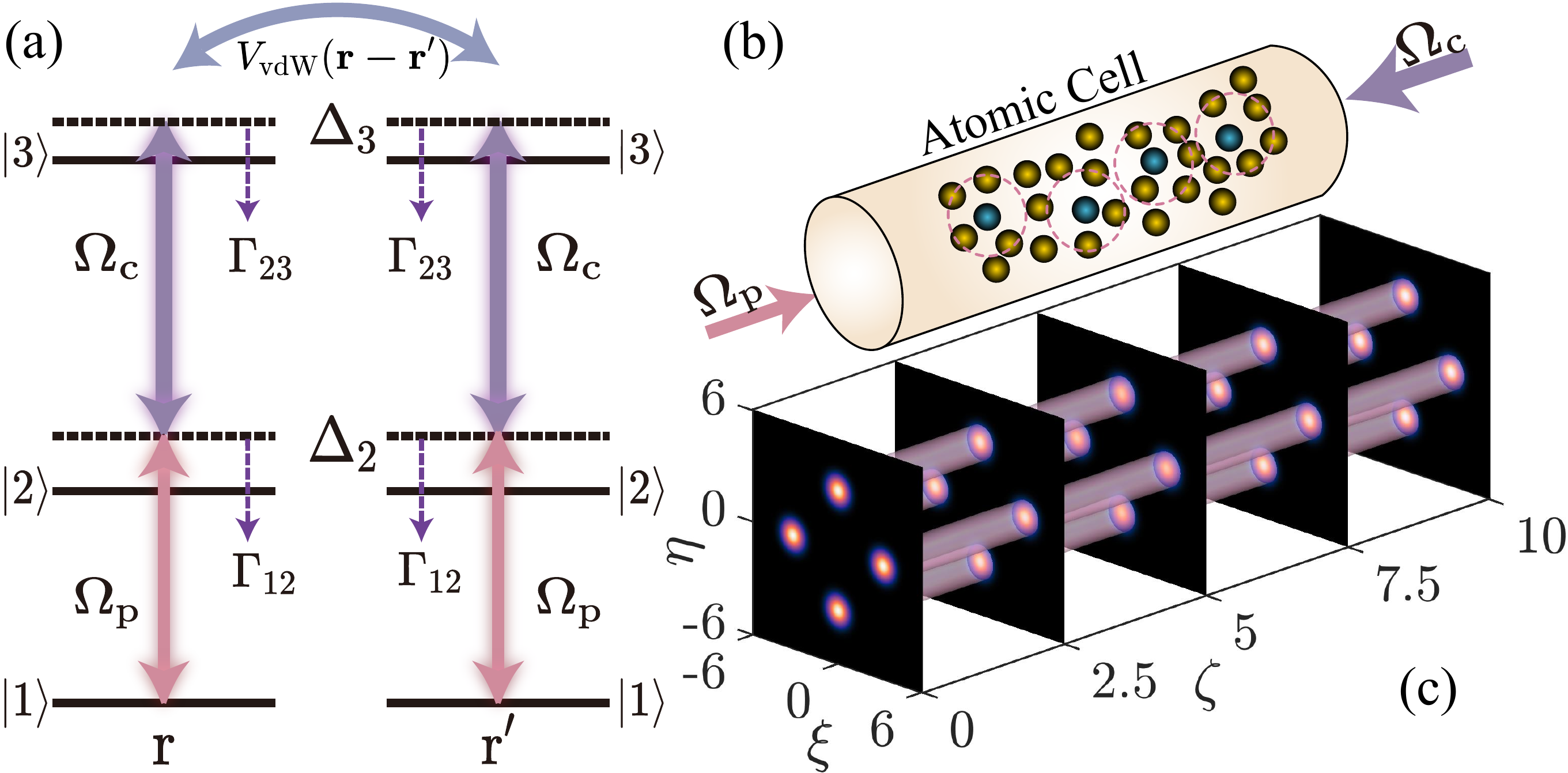}
\caption{(a) The energy-level diagram and excitation scheme for ladder-type three-level atoms. A weak probe optical field (half-Rabi frequency $\Omega_p$) drives the transition $|1\rangle \leftrightarrow |2\rangle$. A strong control laser field (half-Rabi frequency $\Omega_c$) drives the transition $|2\rangle \leftrightarrow |3\rangle$. In the Rydberg state $|3\rangle$, atoms strongly interact through the van der Waals potential $V_{\mathrm{vdW}}(\mathbf{r}'-\mathbf{r})=-\hbar C_6/|\mathbf{r}'-\mathbf{r}|^6$. $\Delta_\alpha$ are the detunings, and $\Gamma_{\alpha\beta}$ $(\alpha<\beta)$ are the spontaneous-emission decay rates. (b) The long-range interaction between Rydberg atoms blocks the excitation of the atoms within the blockade spheres (indicated by the red dashed circles). In each blocked sphere only one Rydberg atom (a small dark-blue sphere) is excited and other atoms (small dark-yellow spheres) cannot be excited. The red and purple arrows indicates the propagation direction of the probe and control fields. (c) The contactless interaction between four optical solitons, which form a stable $2\times2$ rhombic SM, exhibiting no apparent distortion in the course of the propagation.}
\label{fig1}
\end{figure}

Using the Rydberg nonlocal nonlinearity, it has been shown that SMs and
vortex molecules (bound states of solitary vortices) can be
generated and stored in the Rydberg atomic gas~\cite{qin2022stable}.
However, the underlying mechanisms of the formation of SMs and breather
molecules (BMs) in these settings have not yet been fully understood. In
particular, a remaining open question is how to design various SM and BM
states in such systems. With its easily adjustable parameters, the Rydberg
atomic medium allows the nonlocality ($\sigma \equiv R_{b}/R_{0}$) to be
continuously tuned from the local limit to the strongly nonlocal one, $R_{b}$
and $R_{0}$ being the characteristic length of the Rydberg nonlinearity and
beam width, respectively.

In this work, we address mechanisms governing the
formation of stable (2+1)D optical SMs and BMs for different nonlocality
degrees. Specifically, we examine how the initial velocity of circular
motion, which may be imposed onto the Rydberg-EIT medium, affects the resulting
rotation period of SMs and breathing period of BMs.
% It was also observed that the soliton molecule generation mechanisms differ in the nonlocal and strongly nonlocal regimes.
% In the nonlocal regime, SMs arise from intrinsic long-range nonlocal interactions~\cite{qin2022stable}, whereas in the strongly nonlocal regime the response function effectively reduces to a ``harmonic potential" that also supports a stable soliton cluster~\cite{song2018controllable,song_spiraling_2019}.
% {\color{blue}Unlike the existing literature~\cite{qin2022stable}, our study systematically elaborates, based on theoretical predictions, the formation mechanisms of SMs under different degrees of nonlocality in Rydberg-EIT systems. It presents a wider variety of SM configurations and employs a multi-level approach from analytical calculations to numerical simulations to deeply investigate the impact of key parameters such as initial velocity and incident power on the formation of both SMs and BMs, as well as their dynamical behaviors.}
% {\color{blue} Using the Rydberg nonlocal nonlinearity, it has been shown that SMs and vectrox molecules can be generated in the Rydberg atom gas~\cite{qin2022stable}.} However, in many aspects this work is different from Ref.~\cite{qin2022stable}. First,
We find that the giant nonlocal optical nonlinearity supports a variety of
SM configurations, such as rectangular and oblique rhombic ones,
checkerboard cells, and hexagons. They feature a large size and can be
efficiently manipulated by tuning the nonlocality degree of the
nonlinearity. We analyze the binding energy (BE) of the multisoliton
interaction, proceeding from weak to strong nonlocality. Stable SMs can be
formed by long-range interactions between quiescent solitons. On the other
hand, in the strongly nonlocal regime, the SMs are supported by the balance
of strong attractive interactions and the centrifugal force induced by the
orbital motion of the interacting solitons, the rotation period being
controllable by adjusting the system's parameters. Furthermore, the
rotational motion maintains periodic centrifugal and centripetal
\textquotedblleft breathings\textquotedblright\ of the solitons, which may
bind themselves into BMs. By tuning the initial velocity of the solitons,
one can manipulate the evolution of the SMs and BMs and realize their mutual
conversion. The numerical results, which agree very well with analytical
predictions, reveal roles played by the nonlocality degree and initial
 velocity in the generation of the SMs and BMs, and provide a
scenario for manipulating their size, rotation period, and breathing
characteristics. Thus, the present study offers a new scheme for engineering
diverse SM species, that may find applications in data processing and
transmission in optical systems.

\section{The model and light-propagation equations}

\label{sec2}

\subsection{The physical model}

We start by considering a laser-cooled, dilute three-level atomic gas interacting with the combination of a weak probe optical field, driving the transition $|1\rangle \leftrightarrow |2\rangle$ (with half-Rabi frequency $\Omega_p$ and central frequency $\omega_p$), and a strong control laser field, driving the transition $|2\rangle \leftrightarrow |3\rangle$ (with half-Rabi frequency $\Omega_c$ and central frequency $\omega_c$); see Fig.~1(a). Detunings $\Delta_\alpha$ ($\alpha = 2, 3$) define the difference between the laser frequency and atomic transitions, while $\Gamma_{\alpha\beta}$ are rates of the spontaneous emission decay $|\beta\rangle \to |\alpha\rangle$.

In this setting, level $|3\rangle $ represents a high-lying Rydberg state.
The interaction between two Rydberg atoms, located at positions $\mathbf{r}$
and $\mathbf{r^{\prime }}$, is accounted for by the van der Waals potential $%
V_{\mathrm{vdW}}=-\hbar C_{6}/|\mathbf{r^{\prime }}-\mathbf{r}|^{6}$, with
the dispersion coefficient~$C_{6}$ \cite{Saffman2010}. When light propagates
in the medium, the Rydberg excitation in a vicinity of a Rydberg atom is
strongly suppressed, due to the long-range Rydberg-Rydberg interaction;
see Fig.~\ref{fig1}(b). The Rydberg blockade leads to nonlocal nonlinear
optically-mediated interactions~\cite{murray2016quantum}. Note that, in the
excitation scheme shown in Fig.~\ref{fig1}(a), the transition $|1\rangle
\rightarrow |2\rangle \rightarrow |3\rangle $ forms a ladder-shaped
Rydberg-EIT.

Under the electric-dipole and rotating-wave approximations, the Hamiltonian
of the system is $\hat{H}=\mathcal{N}_{\mathrm{a}}\int d^{3}\mathbf{r}\,\hat{%
\mathcal{H}}$ ($d^{3}\mathbf{r}=dxdydz$) with the Hamiltonian density
\begin{align}
\hat{\mathcal{H}}=-& \sum_{\alpha =2}^{3}\hbar \Delta _{\alpha }\hat{S}%
_{\alpha \alpha }(\mathbf{r},t)-\hbar \left[ \Omega _{\mathrm{p}}\hat{S}%
_{12}+\Omega _{\mathrm{c}}\hat{S}_{23}+\mathrm{h.c.}\right]  \notag \\
+& \mathcal{N}_{\mathrm{a}}\int d^{3}\mathbf{r}^{\prime }\hat{S}_{33}(%
\mathbf{r}^{\prime },t)V_{\mathrm{vdW}}(\mathbf{r}^{\prime }-\mathbf{r})\hat{%
S}_{33}(\mathbf{r},t),  \notag  \label{H}
\end{align}%
where $\mathcal{N}_{\mathrm{a}}$ is the atomic density, $\Delta _{2}$ and $%
\Delta _{3}$ are, respectively, the one- and two-photon detunings; $\hat{S}%
_{\alpha \beta }\equiv |\beta \rangle \langle \alpha |\exp \{i[(\mathbf{k}%
_{\beta }-\mathbf{k}_{\alpha })\cdot \mathbf{r}-(\omega _{\beta }-\omega
_{\alpha }+\Delta _{\beta }-\Delta _{\alpha })t]\}$ is the operator of the
atomic transition between $|\alpha \rangle $ and $|\beta \rangle $. Further,
$\Omega _{\mathrm{p}}=(\mathbf{e}_{\mathrm{p}}\cdot \mathbf{p}_{12})\mathcal{%
E}_{\mathrm{p}}/(2\hbar )$ and $\Omega _{\mathrm{c}}=(\mathbf{e}_{\mathrm{c}%
}\cdot \mathbf{p}_{23})\mathcal{E}_{\mathrm{c}}/(2\hbar )$ represent the
Rabi half-frequencies for the probe and control laser fields, respectively,
with $\mathbf{p}_{\alpha \beta }$ being the electric-dipole matrix element
associated with the transition $|\beta \rangle \leftrightarrow |\alpha
\rangle $.

The atomic dynamics is governed by the Heisenberg equations of motion for
operators $\hat{S}_{\alpha \beta }(\mathbf{r},t)$, i.e., $i\hbar {\partial }%
_{t}\hat{S}_{\alpha \beta }(\mathbf{r},t)=[\hat{S}_{\alpha \beta }(\mathbf{r}%
,t),\hat{H}]$. Taking the expectation values on both sides of this equation,
we obtain the expectation-value equations for the operator, i.e., $\langle
\hat{S}_{\alpha \beta }(\mathbf{r},t)\rangle $.
% the density matrix equation with matrix elements $\rho_{\alpha\beta}(\mathbf{r},t) \equiv \langle \hat{S}_{\alpha\beta}(\mathbf{r},t) \rangle$.
To include the decay of the atomic levels due to the spontaneous emission,
relaxation constants $\Gamma $ associated with each ${\rho }_{\alpha \beta }$
are introduced~\cite{Scully1997}. Using the definitions $\rho _{\alpha \beta
}(\mathbf{r},t)\equiv \langle \hat{S}_{\alpha \beta }(\mathbf{r},t)\rangle $%
, the dynamics of the density matrix $\hat{\rho}$ is governed by the Bloch
equation,
\begin{equation}
\begin{aligned} \frac{\partial \hat{\rho}}{\partial t} = -\frac{i}{\hbar}
\left[ \hat{H}, \hat{\rho} \right] + \Gamma \left[ \hat{\rho} \right],
\label{bloc} \end{aligned}
\end{equation}%
where $\hat{\rho}(\mathbf{r},t)$ is a $3\times 3$ density matrix, with
elements $\rho _{\alpha \beta }$, $\alpha ,\,\beta =1,\,2,\,3$, that governs
the evolution of the atomic population and coherence. The explicit form of Eq.~(\ref{bloc}) is presented in Appendix~\ref{AP:BE}.

The propagation of the probe field is governed by the Maxwell equation.
Under the paraxial and slowly-varying-envelope approximations, it is reduced
to
\begin{equation}
i\left( \frac{\partial }{\partial z}+\frac{1}{c}\frac{\partial }{\partial t}%
\right) \Omega _{\mathrm{p}}+\frac{c}{2\omega _{p}}\nabla _{\perp
}^{2}\Omega _{\mathrm{p}}+\kappa _{12}\rho _{21}=0,  \label{MB}
\end{equation}%
where $\nabla _{\perp }^{2}=\partial _{x}^{2}+\partial _{y}^{2}$, and $%
\kappa _{12}=\mathcal{N}_{\mathrm{a}}\omega _{p}|\mathbf{p}%
_{12}|^{2}/(2\varepsilon _{0}c\hbar )$, with $\omega _{p}$, $\varepsilon
_{0} $, and $c$ being, respectively, the central frequency of the weak probe
laser field, vacuum dielectric constant, and speed of light in vacuum.

\subsection{The nonlocal nonlinear Schr\"{o}dinger (NLS) equation}

As the probe is much weaker than the control field, the system of the
Maxwell-Bloch (MB) equations~(\ref{bloc}) and (\ref{MB}) can be solved by
means of the perturbation theory. Here, we are interested in the
steady-state properties of the system, for which the time derivative in the
MB equations may be neglected, which is valid for the probe field with a
long temporal duration. On the other hand, the Rydberg-Rydberg interaction
is treated beyond  mean-field approximation~\cite{bai2019stable}.
Thus, we solve the MB equations up to the third order with respect to $%
\Omega _{p}$. The result is the (2+1)D nonlocal NLS equation for the probe
field \cite{qin2022stable,bai2019stable},
\begin{align}
&i\frac{\partial }{\partial z}\Omega _{\mathrm{p}}+\frac{c}{2\omega _{\mathrm{%
p}}}\nabla _{\perp }^{2}\Omega _{\mathrm{p}}+W\left\vert \Omega _{\mathrm{p}%
}\right\vert ^{2}\Omega _{\mathrm{p}}   \notag \\
&~+\int {d^{2}{\mathbf{r}_{\perp }^{\prime }}G(\mathbf{r}_{\perp },\mathbf{%
r_{\perp }^{\prime }}){{\left\vert {{\Omega _{\mathrm{p}}(\mathbf{r_{\perp
}^{\prime }},z)}}\right\vert }^{2}}{\Omega _{\mathrm{p}}(\mathbf{r}_{\perp
},z)}}=0,  \label{nls}
\end{align}%
with $\mathbf{r}_{\perp }=(x,y)$ and $d^{2}\mathbf{r}_{\perp }^{\prime
}=dx^{\prime }dy^{\prime }$. The last two terms in Eq.~(\ref{nls}) are
contributed by the short-range local and long-range nonlocal optical Kerr
nonlinearities, respectively.
Explicit expressions of $W$ and $G$
% respectively come from the local nonlinear and the nonlocal nonlinear
% responseand respectively,
are given in Appendix~\ref{AP:WG}.

We convert the
propagation equation~(\ref{nls}) into a dimensionless form,
\begin{align}
&i\frac{\partial u}{\partial \zeta }+\left( \frac{\partial ^{2}}{\partial \xi
^{2}}+\frac{\partial ^{2}}{\partial \eta ^{2}}\right) u+w|u|^{2}u  \notag \\
&~+u\iint d\xi ^{\prime }d\eta ^{\prime }g(\xi ^{\prime }-\xi ,\eta ^{\prime
}-\eta )\left\vert u\left( \xi ^{\prime },\eta ^{\prime }\right) \right\vert
^{2}=0,  \label{nnls}
\end{align}%
where $u=\Omega _{\mathrm{p}}/U_{0}$, $\zeta =z/(2L_{\mathrm{diff}})$, $(\xi
,\eta )=(x,y)/R_{0}$, $w=2L_{\mathrm{diff}}|U_{0}|^{2}W$, and $g(\xi
^{\prime }-\xi ,\eta ^{\prime }-\eta )=2L_{\mathrm{diff}%
}R_{0}^{2}|U_{0}|^{2}G(\xi ^{\prime }-\xi ,\eta ^{\prime }-\eta )$, with the
diffraction length given by $L_{\mathrm{diff}}=\omega _{p}R_{0}^{2}/c$.
Here, $U_{0}$ and $R_{0}$ are typical Rabi half-frequency and beam radius,
respectively.

To address a typical example, we refer to $^{88}\mathrm{Sr}$ atoms with
atomic levels $|1\rangle =|5s^{2}\,^{1}S_{0}\rangle $, $|2\rangle
=|5s5p\,^{1}P_{1}\rangle $, and $|3\rangle =|5sns\,^{1}S_{0}\rangle $. For
the principal quantum number $n=60$, the dispersion parameter is $%
C_{6}\approx 2\pi \times 81.6\,\mathrm{GHz}\,\mathrm{\mu }\mathrm{m}^{6}$
\cite{robicheaux2019calculations,guoDipolePolarizabilitiesMagic2010}. As $%
C_{6}>0$, the Rydberg-Rydberg interaction is attractive, making it possible
to support stable bright multi-dimensional solitons~\cite{tikhonenkov2008anisotropic,qin2022stable}.
The spontaneous emission decay rates and detunings are, respectively, $%
\Gamma _{12}\approx 2\pi \times 32$ MHz, $\Gamma _{23}\approx 2\pi \times
16.7$ kHz, and $\Delta _{2}=-2\pi \times 240$ MHz , $\Delta _{3}=2\pi \times
0.4$ MHz. The relevant value of the density of the atomic gas is $\mathcal{N}%
_{\mathrm{a}}=2\times 10^{11}$ cm$^{-3}$, and the half-Rabi frequency of the
control field is $\Omega _{\mathrm{c}}=2\pi \times 9$ MHz. The obvious
condition, $\Delta _{2}\gg \Gamma _{12}$, places the system in the
dispersive-nonlinearity regime, in which the imaginary parts of coefficients
in Eq.~(\ref{nls}) are much smaller than their real counterparts. Therefore,
Eqs.~(\ref{nls}) and (\ref{nnls}) are written, in the first approximation,
with real coefficients~\cite{bai2019stable}.

As mentioned above, the exact form of the nonlocal nonlinear response
function $g(\xi ^{\prime }-\xi ,\eta ^{\prime }-\eta )$ is very cumbersome,
it may be represented, with a reasonable accuracy, by a Gaussian~\cite%
{qin2022stable,Hang2019},
\begin{equation}
g\approx \frac{g_{0}}{\pi (0.93\sigma )^{2}}\exp \left[ -\frac{\left( \xi
-\xi ^{\prime }\right) ^{2}+\left( \eta -\eta ^{\prime }\right) ^{2}}{%
(0.93\sigma )^{2}}\right] ,  \label{sigma}
\end{equation}%
where $g_{0}=\iint d\xi ^{\prime }d\eta ^{\prime }\,g(\xi ^{\prime }-\xi
,\eta ^{\prime }-\eta )$ is a constant, and $\sigma \equiv R_{\mathrm{b}%
}/R_{0}$ characterizes the nonlocality degree of the nonlinearity, with $R_{%
\mathrm{b}}=(|C_{6}/\delta _{\mathrm{EIT}}|)^{1/6}$ denoting the
Rydberg-blockade radius, which is determined by the linewidth of the EIT
transmission spectrum (for $|\Delta _{2}|\gg \Gamma _{12}$), $\delta _{%
\mathrm{EIT}}\approx |\Omega _{\mathrm{c}}|^{2}/|\Delta _{2}|$. With the
parameters adopted above, we get $R_{\mathrm{b}}\approx 7\,\mathrm{\mu }$%
m.

According to the degree of the nonlocality, the following analysis is
reported in three distinct regimes \cite{PhysRevE.64.016612,bai2019stable}:
(i) the local response, $\sigma \ll 1$; (ii) the nonlocal response, $\sigma
\sim 1$; (iii) the strongly nonlocal response, $\sigma \gg 1$.

%===========================fig2===============================%
\begin{figure*}[ht]
\centering
\includegraphics[width=\linewidth]{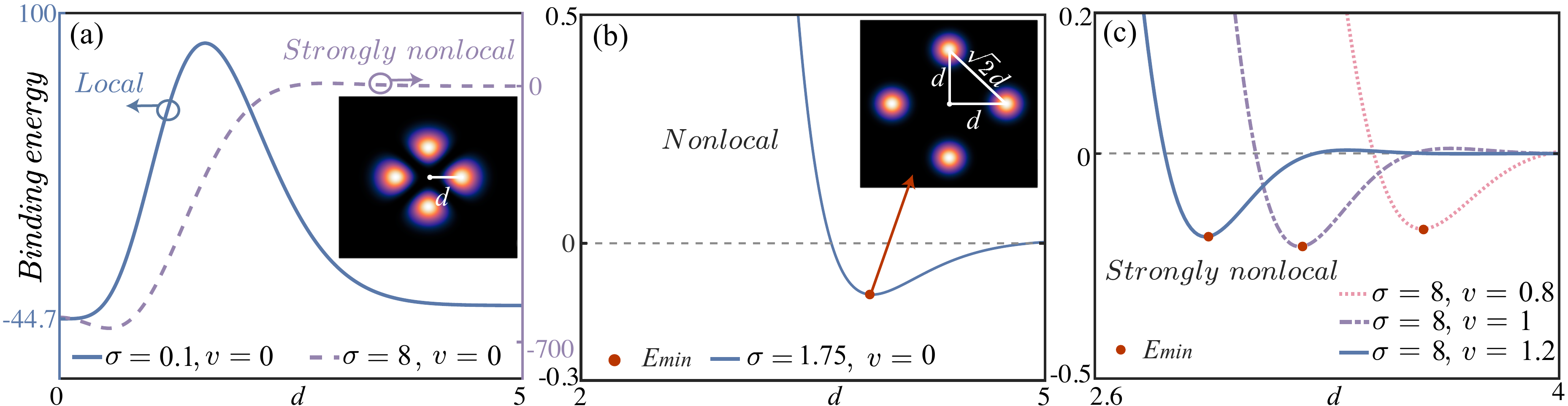}
\caption{{\protect\footnotesize The binding energy (BE) of the four-soliton
rectangular rhombic chain (\protect\ref{rhombus2}). (a) In the case of
nearly local nonlinearity [$\protect\sigma =0.1$ in Eq. (\protect\ref{sigma}%
)], the BE has no minimum, indicating the absence of stable SMs. In the case
of strongly nonlocal nonlinearity ($\protect\sigma =8$), a shallow minimum
is observed at $d=0.61$, suggesting the formation of a small-size
non-rotating SM. The parameters are $A=4$, $a=1$, $b=0$, and $v=0$. (b) In
the regime of moderately nonlocal nonlinearity ($\protect\sigma =1.75$), a
stable non-rotating SM, shown in the inset, is found at $d=3.87$. The
parameters are: $A=4$, $a=0.95$, and $v=0$. (c) In the regime of strongly
nonlocal nonlinearity ($\protect\sigma =8$), varying the initial rotation
velocity leads to stable rotating rhombic SMs with different sizes: $d=3.6$
for $v=0.8$; $d=3.23$ for $v=1$; and $d=2.94$ for $v=1.2$. Other parameters
are $A=4$, $a=1$, and $b=0$. }}
\label{fig2}
\end{figure*}

\section{Nonlocal (2+1)D SMs}

\label{sec3}

\subsection{The binding energy (BE)}

To investigate the stability of the SMs, we aim to calculate the BE between $%
N$ widely
separated spatial solitons (i.e., ones involved in the interaction of the \textit{noncontact type}), 
approximated by the superposition
\begin{equation}
u_{m}(\xi ,\eta )=\sum_{n=1}^{N}u_{n}(\xi -\xi _{n},\eta -\eta _{n}),
\label{Eq:ansatz}
\end{equation}%
composed of $N$ Gaussians, each one, with amplitude $A$, width $a$, and
chirp (phase-front curvature) $b$~ \cite{desaix1991variational}, adopted as
\begin{align}
u_{n}=&A\exp \Bigl\{ -\frac{(\xi -\xi _{n})^{2}+(\eta -\eta _{n})^{2}}{2a^{2}}%
+ib(\xi ^{2}+\eta ^{2})\notag\\
    &+i[\left( v_{\xi }\right) _{n}(\xi -\xi _{n})+\left(
v_{\eta }\right) _{n}(\eta -\eta _{n})]+i\theta _{n}\Bigr\}.  \label{Gauss}
\end{align}%
The phase shift between
adjacent solitons is $\theta _{n+1}-\theta _{n}=\pi $, making the contact
interaction between them repulsive, which is necessary for the stability of
the appearing SMs \cite{qin2022stable,qin2023}. The initial speeds of the soliton cluster (\ref{Gauss}), set in rotational
motion with angular velocity $v$, are $\left\{ \left( v_{\xi }\right)
_{n},\left( v_{\eta }\right) _{n}\right\} =v\cdot \left\{ -\eta _{n},\xi
_{n}\right\} $.

The simplest structure under the consideration is a rectangular rhombic SM ($%
N=4$), which is initially composed of four solitons, placed at positions
\begin{align}
&(\xi _{1},\eta _{1})=(d_{\xi },0),~~~(\xi _{2},\eta _{2})=(0,d_{\eta }),\notag\\
&(\xi_{3},\eta _{3})=(-d_{\xi },0),~(\xi _{4},\eta _{4})=(0,-d_{\eta }),
\label{rhombus2}
\end{align}
where $d_{\xi}$ and $d_{\eta}$ are the distance from the origin of the coordinate system. A more
general SM\ configuration represents an oblique (compressed) rhombus,
% with vertices at points
with $d_{\xi }\neq d_{\eta }$, see Fig. \ref{fig4} below.

BE is the difference between the SM energy of a molecule [currently
represented by ansatz (\ref{Eq:ansatz})], and the sum of the energies of
isolated individual solitons bound in the molecule, i.e., $E_{\mathrm{bind}%
}=E[u_{m}]-\sum_{n=1}^{N}E[u_{n}]$~\cite%
{alkhawajaInteractionForcesTwodimensional2012}. It is computed as per the
energy functional corresponding to Eq. (\ref{nnls}):
\begin{align}
E[u]=& \iint d\xi d\eta \left[ \left\vert \frac{\partial u}{\partial \xi }%
\right\vert ^{2}+\left\vert \frac{\partial u}{\partial \eta }\right\vert
^{2}-\frac{w}{2}|u|^{4}\right.   \notag \\
& \left. -\frac{|u|^{2}}{2}\iint d\xi ^{\prime }d\eta ^{\prime }g(\xi
^{\prime }-\xi ,\eta ^{\prime }-\eta )|u|^{2}\right].   \label{E}
\end{align}%
Actually, for the light propagation in the spatial domain, Eq. (\ref{E})
determines not the physical energy, but rather the total Hamiltonian.
The obvious necessary stability condition is the existence of a minimum of $E_{%
\mathrm{bind}}$.

Along with energy (\ref{E}), the underlying NLS equation (\ref{nnls})
conserves the total norm, alias the scaled integral power of the optical
beam, $Q=\iint d\xi d\eta \left\vert u\left( \xi ,\eta \right) \right\vert
^{2}$. Dealing below with rotating modes, it is relevant to consider the
angular momentum, which is a dynamical invariant too, $M=i\iint d\xi d\eta
u^{\ast }\left( \xi\partial _{\eta}u-\eta\partial _{\xi}u\right) $~\cite{desyatnikovRotatingOpticalSoliton2002a}, where $\ast $
stands for the complex conjugation.

In the case of weak nonlocal nonlinearity, it can be approximated by the local limit,
\textit{viz}., $g(\xi -\xi ^{\prime },\eta -\eta ^{\prime })\rightarrow
g_{0}\delta (\xi -\xi ^{\prime },\eta -\eta ^{\prime })$, with $g_{0}=\iint
d\xi ^{\prime }d\eta ^{\prime }g(\xi ^{\prime },\eta ^{\prime })$, which
converts Eq.~(\ref{nnls}) into local NLS equation. In
Fig.~\ref{fig2}(a), the solid line presents the BE as a function the half-size
$d=d_{\xi}=d_{\eta}$ of the rectangular rhombus (\ref{rhombus2}), with other parameters fixed
as $\sigma =0.1$, $A=4$, and $a=1$. In this case, no local minimum of $E_{%
\mathrm{bind}}$ is identified, suggesting that a stable bound state cannot
form under the present conditions.

In the limit of strong nonlocality, represented, e.g., by $\sigma =8$ in Eq.
(\ref{sigma}), the nonlinear terms in Eq.~(\ref{nnls}) can be approximated
by a quasilinear expression, similar to that in the model of
\textquotedblleft accessible solitons" \cite{snyderAccessibleSolitons1997}
% $gP_0\,u$, where $P_0=\iint d^2\zeta |u|^2$ is the power of the probe field.
$g_{1}u+g_{2}(\xi ^{2}+\eta ^{2})u$, with constants $g_{1}$ and $g_{2}$, see
Eq. (\ref{gg}) in Appendix~\ref{AP:A}. In this regime, the BE is plotted vs.
$d$ by the dashed line in Fig.~\ref{fig2}(a), featuring a shallow minimum,
i.e., a possible bound state of the \textit{contact type}, at a small half-size $d\approx 0.61$, the
respective separation-to-width ratio being $s\equiv \sqrt{2}d/a\approx 0.86$
[recall the soliton's width $a$ is defined by the Gaussian ansatz (\ref{Gauss}%
)]. This result implies that the repulsive force, provided by the
out-of-phase setup of the four-soliton chain, is actually too weak to
balance the strong effective quadratic potential, $g_{2}(\xi ^{2}+\eta ^{2})$%
, thereby hindering the creation of an SM with a reasonably larger size.

Below, our analysis focuses on large-size SMs achievable when the
nonlocality degree $\sigma $ takes intermediate values. For instance, the
previous work \cite{qin2022stable} has shown that $\sigma =1.75$ can
stabilize the quiescent (non-rotating) rhombic SM. Figure~\ref{fig2}(b)
illustrates this case, with the inset showing a specific example of the
rhombic SM with $A=4$, $a=0.95$, $d=3.87$, $\sigma =1.75$, and $v=0$.

\subsection{Rectangular rhombic SMs}

We proceed with the search for stable (2+1)D spatial SMs numerically solving Eq.~(\ref{nnls}) by means of the split-step Fourier method~\cite{YangJK}.
% {\LARGE [It is necessary to
% briefly outline the numerical methods used in the work, and exactly specify
% the size of the domain and boundary conditions.]}
The initial conditions for
the simulations are set to match those presented in Fig.~\ref{fig2}(b), with a
small random perturbation, represented by factor $[1+\epsilon R(\xi ,\eta )]$
multiplying the input. Here, $\epsilon \ll 1$ is the perturbation amplitude,
and $R$ is a random variable uniformly distributed in the interval $[-1,+1]$.

Owing to the transverse instability in the case of the nearly local
nonlinearity ($\sigma =0.1$), the multisoliton propagation shown in Fig.~\ref%
{fig3}(a) is highly unstable. This simulation was initiated by
expressions (\ref{Eq:ansatz}), (\ref{Gauss}), and (\ref{rhombus2}) with
parameters $A=4$, $a=0.95$, $d_{\xi}=d_{\eta}=d=3.87$, $b=0$, $Q=181.4$, and perturbation amplitude $%
\epsilon =0.05$.

%===========================fig3===============================%
\begin{figure}[b]
\centering
\includegraphics[width=1\columnwidth]{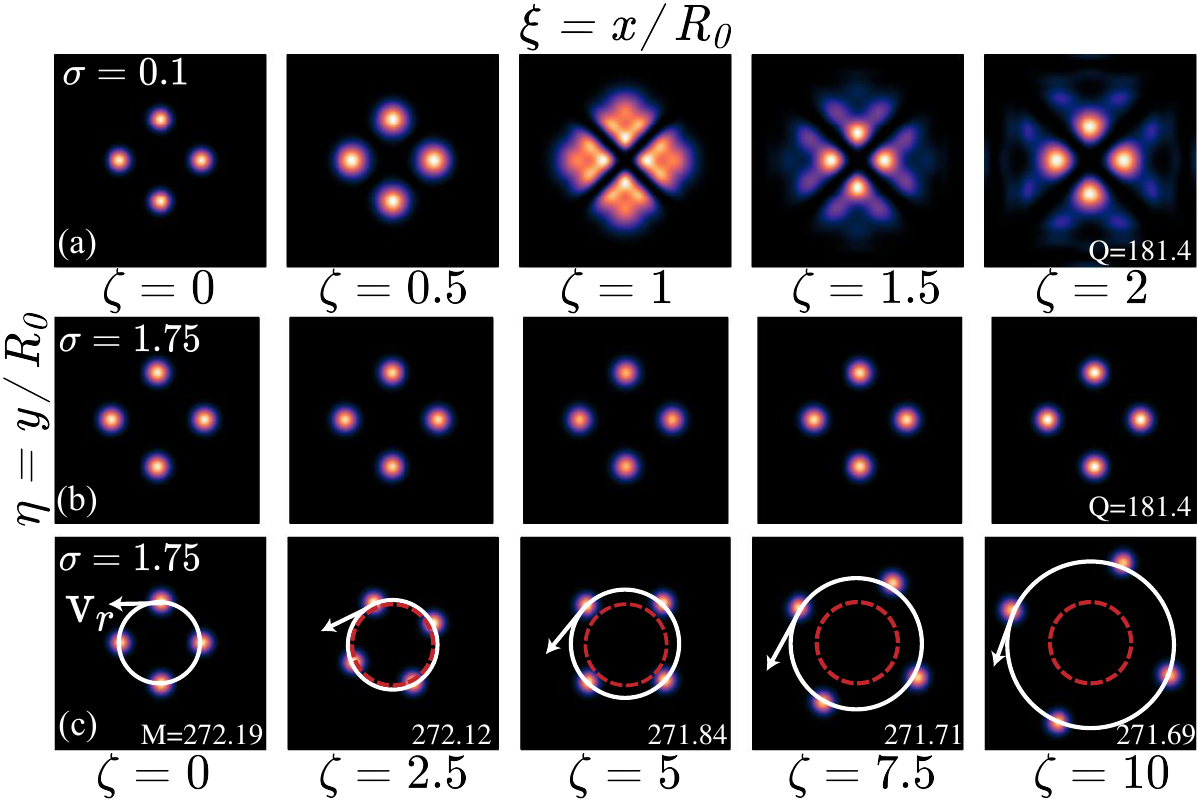}
\caption{{\protect\footnotesize The simulated evolution of rhombic SMs. The
initial conditions are set by Eqs. (\protect\ref{Eq:ansatz}), (\protect\ref%
{Gauss}), and (\protect\ref{rhombus2}), with $A=4$, $a=0.95$, $d=3.87$, $b=0
$, and a }${\protect\footnotesize 5\%}${\protect\footnotesize \ random
perturbation added to the input. (a)~In the case of the nearly local
nonlinearity, with $\protect\sigma =0.1$ in Eq. (\protect\ref{sigma}), the
solitons with power $Q=181.4$ are unstable. (b)~ In the case of moderately nonlocal nonlinearity,
with $\protect\sigma =1.75$, the SM with power $Q=181.4$ is stable. (c)~When the rotation, with
angular speed $v=0.1$, is imposed onto the input, the nonsteadily rotating
SM with a gradually increasing size is produced. The red dashed circle shows
the initial configuration of the four-soliton molecule; while the radius of
the gradually expanding bound state is designated by the white solid curves.
}}
\label{fig3}
\end{figure}
%===========================fig3===============================%
%
Shown in Fig.~\ref{fig3}(b) is the example of the evolution in the case
of moderately nonlocality ($\sigma =1.75$), with the input taken as
per Eq.~(\ref{rhombus2}),
with $d_{\xi}=d_{\eta}=d=3.87$ and $Q=181.4$. The rhombic SM is found to be
stable, as it relaxes to the self-cleaned form, which remains close to the
unperturbed one. A 3D view of the stable propagation is shown in Fig.~\ref%
{fig1}(c). Here, the separation-to-width ratio is $s=\sqrt{2}d/a\approx 5.8$%
. The appearance of such a stable large-size SM is supported by the giant
nonlocal nonlinearity, cf. Ref.~\cite{qin2022stable}.
The maximum average
power density $P_{\mathrm{max}}$,
necessary for the generation of this SM in the
experiment, can be obtained from the magnitude of the corresponding Poynting
vector~\cite{Huang2005}, which produces an estimate $P_{\mathrm{max}}=32$%
~nW. Thus, the power required to generate the stable
SMs is found to be in the nanowatt range, which is, roughly, five orders of
magnitude smaller than the SM-generation power in fiber-laser systems and
optical solid-state waveguides~\cite{Kurtz2020,rotschild2006NP,Wang2019}.

In Fig.~\ref{fig3}(c), the initial rotation with angular velocity $v=0.1$
is applied to the soliton chain initially placed along the red dashed
circle.
As symmetric solitons have angular velocities of equal magnitude and opposite directions, the overall 
rotation velocity of the soliton cluster is zero.
The rotation gives rise to an outward-directed centrifugal force.
As a result, the solitons gradually spiral outward, with their orbital
radius, designated by the solid white curve in Fig.~\ref{fig3}(c),
continuously growing, i.e., a stable rotating SM does not emerge in this
case.
The initial angular velocity pushes the solitons in rotation; in the course of the rotation,
the angular momentum has values $M = 272.19,\,272.12,\,271.84,\,271.71,\,271.69$, corresponding to 
the propagation distances $\zeta=0,\,2.5,\,5,\,7.5,\,10$.

% \subsection{Oblique (compressed) rhombic SMs}

\begin{figure}[t]
\centering\includegraphics[width=1\columnwidth]{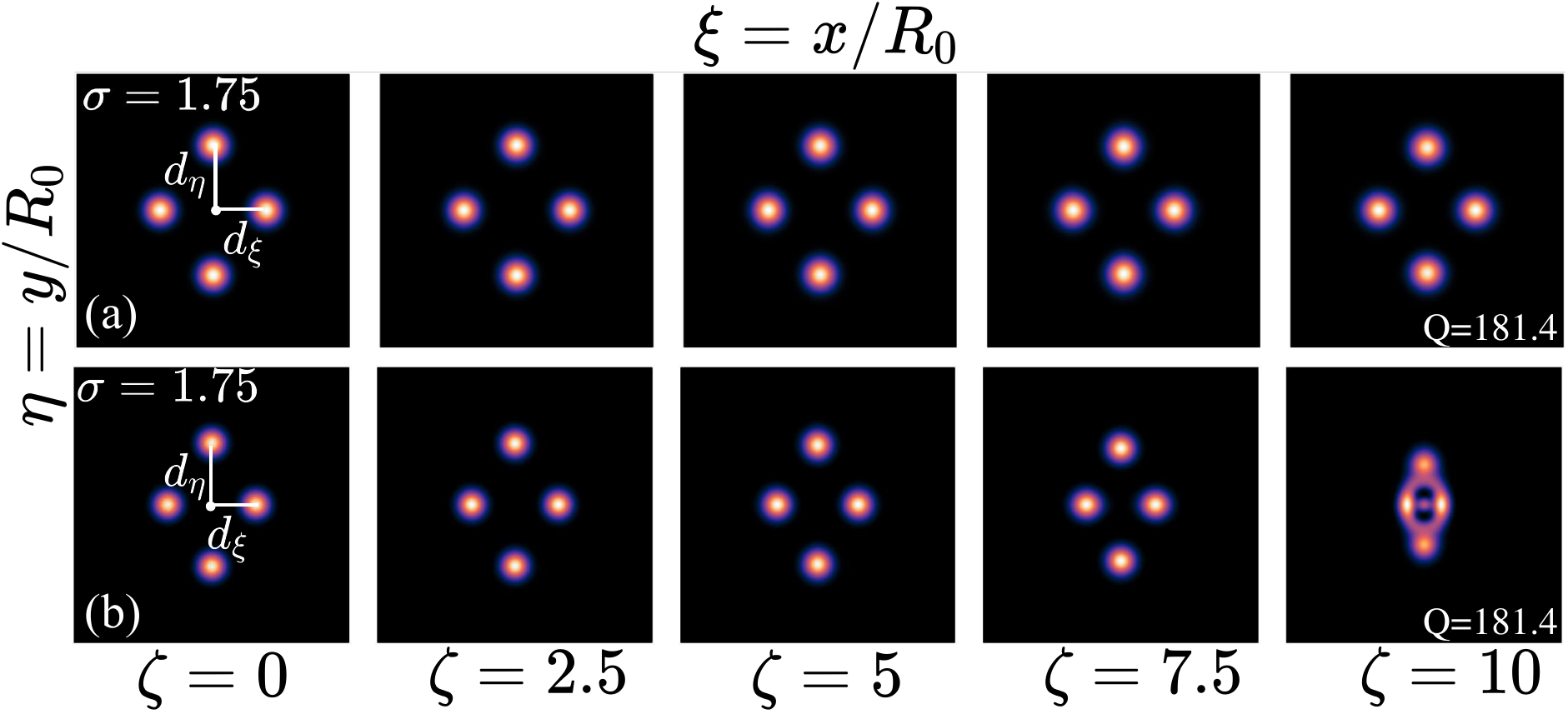}
\caption{{\protect\footnotesize The simulated evolution of oblique rhombic
SMs. (a) The stable SM with power $Q=181.4$, and sizes $d_{\xi }=3.5$, $d_{\eta }=4.27$. (b) The
unstable one with power $Q=181.4$ and size $d_{\xi} =3.3$, $d
_{\eta }=4.47$. All produced by the system with the moderate
nonlocality, $\sigma =1.75$, and parameters $A=4$, $a=0.95$, $b=0$ in the
Gaussian input (\protect\ref{Gauss}).}}
\label{fig4}
\end{figure}

In addition to the rectangular rhombuses, stable four-soliton molecules can
be constructed in the form of non-rotating rhombuses with unequal sizes in
the horizontal and vertical directions $d_{\xi }=3.5$ and $d_{\eta }=4.27$, as shown
in Fig. \ref{fig4}(a), for the compression factor $d_{\xi
}/d_{\eta }\approx \allowbreak 0.819$ and power $Q=181.4$. On the other hand, Fig. \ref{fig4}(b) shows that a still stronger compression, to $d_{\xi
}/d_{\eta }\approx \allowbreak 0.716$ with $d_{\xi }=3.3$ and $d_{\eta }=4.47$, leads to the loss of stability of the
oblique rhombus. Thus, there is a stability boundary for the compression
factor of the oblique rhombuses between these two values, for the same
moderate nonlocality degree, $\sigma =1.75$, and the same parameters of
Gaussian (\ref{Gauss}), $A=4$, $a=0.95$, $b=0$, as in Fig. \ref{fig3}(b).
%%%%%%%%%%%%%%%%%%%%%%%%%%%%%%%%%%%%%%%%%%

\subsection{Multi-soliton bound states}

\begin{figure}[t]
\centering
\includegraphics[width=1\columnwidth]{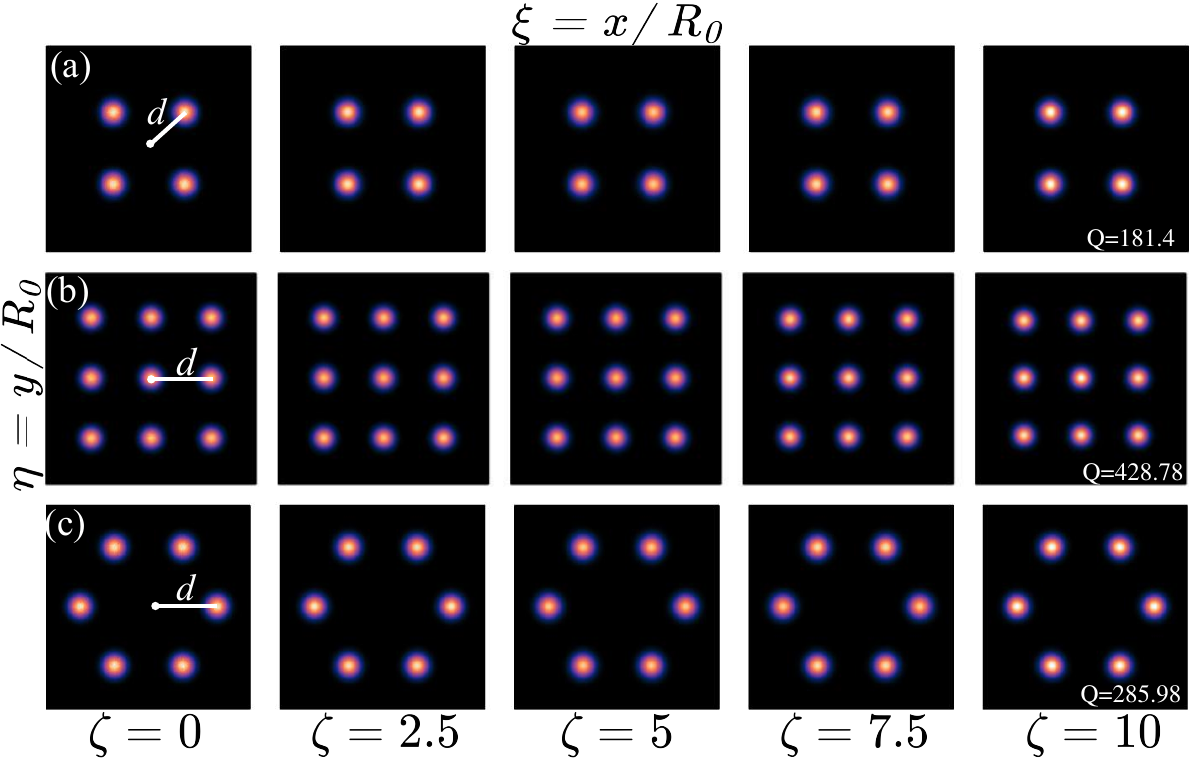}
\caption{{\protect\footnotesize The evolution of several species of stable
SMs in the regime of the moderate strength of the nonlinear nonlocality,
with $\protect\sigma =1.75$. (a)~The square-shaped SM with size $d=3.87$
and power $Q=181.4$, based on
frame (\protect\ref{square}) . (b)~The }$3\times 3$ {\protect\footnotesize %
checkerboard-cell SM, with size $d=5.68$ and power $Q=428.78$. (c)~The hexagonal SM with size $d=5.57$ and power $Q=285.98$,
based on frame (\protect\ref{hex}). The input parameters in Gaussian (\protect
\ref{Gauss}) are $A=4$, $a=0.95$, and $b=0$.}}
\label{fig5}
\end{figure}

Parallel to the rhombic SMs, the Rydberg-EIT system supports other species
of multi-soliton molecules. As the starting point for producing additional
species, one can rotate a rectangular rhombus by $45$ degrees and consider
it as a square-shaped SM, with the individual solitons set at position%
\begin{equation}
\left( \xi ,\eta \right) =(\pm \sqrt{2}d/2,\pm \sqrt{2}d/2).  \label{square}
\end{equation}%
For the same value $d=3.87$ and $\sigma =1.75$,
as adopted above for the moderate degree of the nonlocality in Fig. \ref{fig3}(b), Fig. \ref{fig5}(a)
demonstrates the stable propagation of the square-shaped SM, with parameters
of Gaussian (\ref{Gauss}) taken as $A=4$, $a=0.95$, $b=0$, and $Q=181.4$. This result
is similar to the stability of the rectangular rhombus in Fig. \ref{fig3}%
(b).

Proceeding from the square cell in Fig. \ref{fig5}(a), it is possible to
produce a new stable SM\ structure, in the form of a checkerboard cell,
composed of $9$ solitons, which is displayed in Fig.~\ref{fig5}(b). The
consideration of the corresponding BE produces the equilibrium value, $%
d=5.68 $, for the same input parameters as in Fig.~\ref{fig5}(a).

%===========================fig4===============================%
%%%%%%%%%%%%%%%%%%%%%%%%%%%%%%%%

A stable SM shaped as a hexagonal cell can be constructed too, as shown in
Fig.~\ref{fig5}(c). In this configuration, the constituent solitons are set as
points
\begin{equation}
(\xi _{n},\eta _{n})=d[\cos (n\pi /3),\sin (n\pi /3)],~n=1,2,3,4,5,6,
\label{hex}
\end{equation}%
and the equilibrium value of the separation, corresponding to the minimum of
the BE, is $d=5.57$,\ for the same set of parameters in the input Gaussian
as used above, \textit{viz}., $A=4$, $a=0.95$, $b=0$, and $Q=258.98$. Thus, we conclude
that the Rydberg-EIT system sustains a variety of stable SM configurations,
including rectangular and oblique rhombic, square-shaped, checkerboard-cell,
and hexagonal ones.

\section{Rotating SMs and BMs (breather molecules) in the regime of strongly nonlocal nonlinearity}

\label{sec4}

\subsection{The effective propagation equation and BE}

We now proceed to the strongly nonlocal regime. Compared to the intensity
profile of the probe field, the response function $G$ in Eq.~(\ref{nls}) is
very flat in a vicinity of $\mathrm{r}_{\perp }^{\prime }=\mathrm{r}_{\perp
} $, allowing to use its Taylor expansion, $G(\mathrm{r}_{\perp }^{\prime }-%
\mathrm{r}_{\perp })\approx G(0)+G^{\prime \prime }(0){\rm r}_{\perp }^{2}/2$~\cite%
{bai2019stable,snyderAccessibleSolitons1997}. In this case, the nonlocal
term in Eq.~(\ref{nnls}) can be reduced to a quasi-linear form, including
the usual 2D harmonic-oscillator potential, so that the simplified equation
becomes
\begin{equation}
i\frac{\partial \Phi }{\partial \zeta }+\left( \frac{\partial ^{2}}{\partial
\xi ^{2}}+\frac{\partial ^{2}}{\partial \eta ^{2}}\right) \Phi +w|\Phi
|^{2}\Phi +g_{2}(\xi ^{2}+\eta ^{2})\Phi =0.  \label{Phi}
\end{equation}%
where $\Phi \equiv u\left( \xi ,\eta ,\zeta \right) \exp (ig_{1}\zeta )$ and
$g_{1}\equiv 2L_{\mathrm{diff}}P_{0}G(0)$, $g_{2}\equiv L_{\mathrm{diff}%
}R_{0}^{2}P_{0}G^{\prime \prime }(0)$. This reduced equation admits an
approximate analytical Gaussian-shaped SM solution, see Appendix~\ref{ap1}.

To study SMs in the strongly nonlocal regime, we refer to a relevant set of
physical parameters: $\Omega _{\mathrm{c}}=2\pi \times 25$ MHz, $\Delta
_{2}=-2\pi \times 260$ MHz, $\Delta _{3}=2\pi \times 13$ MHz, $R_{0}=2.2\,%
\mathrm{\mu }\text{m}$, $\mathcal{N}_{\mathrm{a}}=5.8\times 10^{12}$ cm$^{-3}$, and $C_{6}\approx 2\pi \times 167\,\rm{THz}\cdot\mathrm{\mu}\rm{m}%
^{6}$, with the principal quantum number $n=120$. For these parameters, the
radius of the Rydberg blockade is $R_b=18\,\mathrm{\mu }\text{%
m, and the nonlocality degree is indeed large, }\sigma =R_{b}/R_{0}\approx 8$%
, the other parameters in Eq. (\ref{Phi}) being $w=0.02$ and $g_{2}=-1$.
% With the system’s parameters seted above, we obtain $\chi_{\text{loc}}\approx 10^{-9}$ m$^2$ V$^{-2}$ and $\chi_{\text{nloc}}\approx 10^{-6}$ m$^2$ V$^{-2}$.

The expression for the energy functional corresponding to Eq. (\ref{Phi})
\begin{align}
E[\Phi ]=& \iint d\xi d\eta \left[ \left\vert \frac{\partial \Phi }{\partial
\xi }\right\vert ^{2}+\left\vert \frac{\partial \Phi }{\partial \eta }%
\right\vert ^{2}-\frac{w}{2}|\Phi |^{4}
\right.   \notag \\
& \left.
-\frac{g_{2}}{2}(\xi ^{2}+\eta ^{2})|\Phi |^{2}\right] .  \label{14}
\end{align}%
Analyzing the BE corresponding to the energy functional (\ref{14}), we find
that non-rotating $N$-soliton clusters cannot form large stable SMs in the
strongly nonlocal regime, as shown in Fig.~\ref{fig2}(a) for $N=4$ and is also
valid for larger values of $N$. The analytical result shown in Appendix~\ref{ap1}
clarify this conclusion: Eqs.~\eqref{eq:sub1} and ~\eqref{eq:sub2} with $%
\mathbf{v}_{r}=(\left( v_{\xi }\right) _{n},\left( v_{\eta }\right) _{n})=0
$ predict periodic oscillations of the center-of-mass of each soliton $[\xi
_{n}(\zeta ),\eta _{n}(\zeta )]=[\xi _{n}(0)\cos \alpha ,\eta _{n}(0)\cos
\alpha ]$, preventing the formation of a stationary SM. An alternative
possibility to construct SMs under strong nonlocality is to use the rotation
imposed onto the initial cluster. The respective BE plot in Fig.~\ref{fig9}%
(a) reveals more than one equilibrium point. When the soliton is placed at
the first equilibrium position, the resulting structure is a contact-type
SM, as shown in Fig.~\ref{fig9}(c); details are provided in Appendix~\ref%
{AP:C}. In this work, however, we focus on long-range, non-contact SMs,
therefore only the second equilibrium position is considered in Fig.~\ref{fig2}%
(c).

As seen in Fig.~\ref{fig2}(c) for $\sigma =8$, varying the initial rotation
velocity easily produces large-size SMs. This is possible because the
rotation induces a centrifugal force, enabling the attractive and repulsive
forces to balance each other, and thus form stable SMs. When the initial
velocity is large, the corresponding centrifugal force is strong, making the
separation between the constituent solitons smaller, helping to generate a
stronger attractive force that balances the repulsive centrifugal one. For
example, at $v=1.2$, the equilibrium position is $d=2.94$, see the blue
solid line in Fig.~\ref{fig2}(c). On the other hand, when the initial velocity
is small and the corresponding centrifugal force is weak, hence the
equilibrium separation between the bound solitons increases, reducing the
attractive force. In particular, at $v=0.8$, the equilibrium position is $%
d=3.6$, see the red dashed line in Fig.~\ref{fig2}(c). The parameters of the
Gaussian (\ref{Gauss}) in this case are the same as above, i.e., $A=4$, $a=1$%
, and $b=0$.

%===========================fig===============================%
\begin{figure}[h]
\centering
\includegraphics[width=1\linewidth]{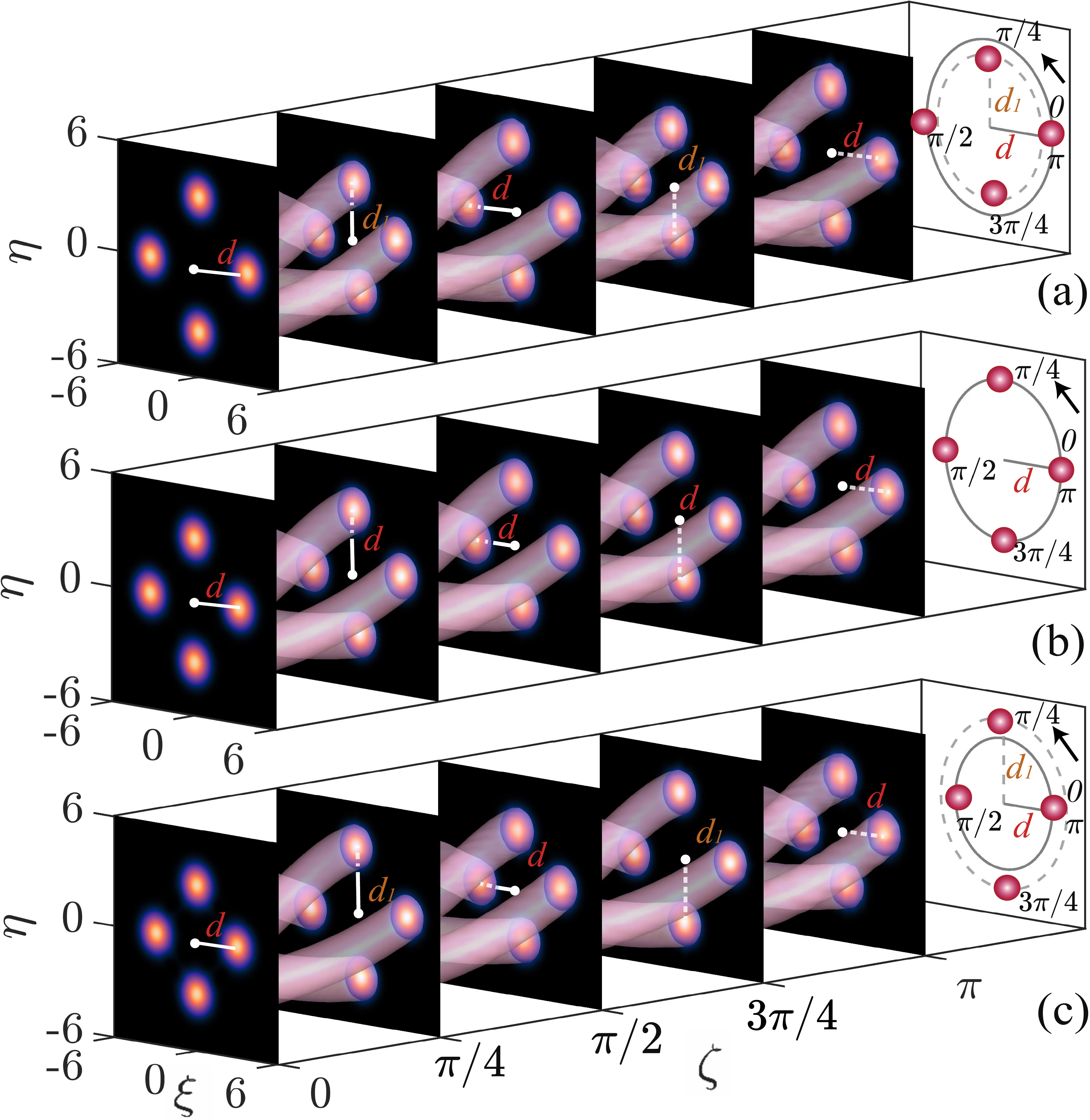}
\caption{{\protect\footnotesize The evolution of rhombic SMs and BMs. (a)
During the propagation, the SM first executes a centripetal motion until it
attains an extremum, then switches to centrifugal motion until it returns to
the initial radius, with initial velocity $v=0.8$. This breathing mode
repeats with the period of $\protect\pi /2$. Simultaneously, the BM
rotates, completing one full turn with period $\protect\pi $. As the initial
rotation velocity is directed counter-clockwise, the BM rotates
counter-clockwise too (as shown by black arrows), with period $\protect\pi $%
. (b) With the initial rotation velocity $v=1$, the SM exhibits
counter-clockwise rotational motion. The radius of the rotation circle is
exactly that of the initial ring. (c) The same as (a), but for $v=1.2$. The
SM first executes centrifugal motion until it attains the maximum, then
switches to centripetal motion until returns to the initial radius. The
rotation and breathing periods are the same as those in panel (a). In the }$%
\left( {\protect\footnotesize \protect\xi ,\protect\eta }\right) $%
{\protect\footnotesize \ cross-sections, $d$ and $d_{1}$ represent the
radius at the corresponding propagation distance $\protect\zeta $. In the
last cross-section, the solid circle marks the initial radial position of
the solitons ($d$); the dashed circle indicates the extrenal radial position
reached during the breathing cycle ($d_{1}$). The four circles mark the
spatial positions of the first soliton at $\protect\zeta =0,\,\protect\pi %
/4,\,\protect\pi /2,$ and $3\protect\pi /4$, respectively. At $\protect\zeta %
=\protect\pi $, the first soliton returns to its initial position. Other
parameters are the same as in Fig.~\protect\ref{fig2}(c).
% Panels (a1), (b1),
% and (c1) display the trajectories corresponding to Figs. (a), (b), and (c),
% respectively: (a1) and (c1) are ellipses with eccentricities 0.60 and 0.55,
% while (b1) is a circle ($e=0$).
}}
\label{fig6}
\end{figure}
%%%%%%%%%%%%%%%%%%%%%%%%%%%%%%%%%%%

\subsection{Rotating SMs and BMs}

Based on the approximate ansatz for the SM solution for far separated
solitons, $\Psi _{N}(\mathbf{r},\zeta )=\sum_{n=1}^{N}\Phi _{n}(\mathbf{r}%
,\zeta )$ (explicit expressions for $\Phi _{n}(\mathbf{r},\zeta )$ are given
in Appendix \ref{ap1}), we derive the centroid equation of motion for the
individual solitons, i.e., $v^{2}\xi _{n}^{2}+\eta _{n}^{2}=v^{2}d^{2}$,
which are actually determined by the Ehrenfest theorem \cite{bohm2012quantum}, see
details in Appendix~\ref{AP:C}. When $v=1$, it reduces
to the circle equation: $\xi _{1}^{2}+\eta _{1}^{2}=d^{2}$, with the
circle's radius exactly equal to that of the initial ring. Therefore, every
constituent Gaussian soliton follows a concentric circular trajectory. When $%
v\neq 1$, the centroid equation of motion reduces to one for the ellipse.
Based on the latter equation, we can define the eccentricity, which is $e=%
\sqrt{1-v^{2}}$ for $0<v<1$, and $e=\sqrt{1-1/v^{2}}$ for $v>1$.

%===========================fig===============================%
\begin{figure}[ht]
\centering
\includegraphics[width=0.95\linewidth]{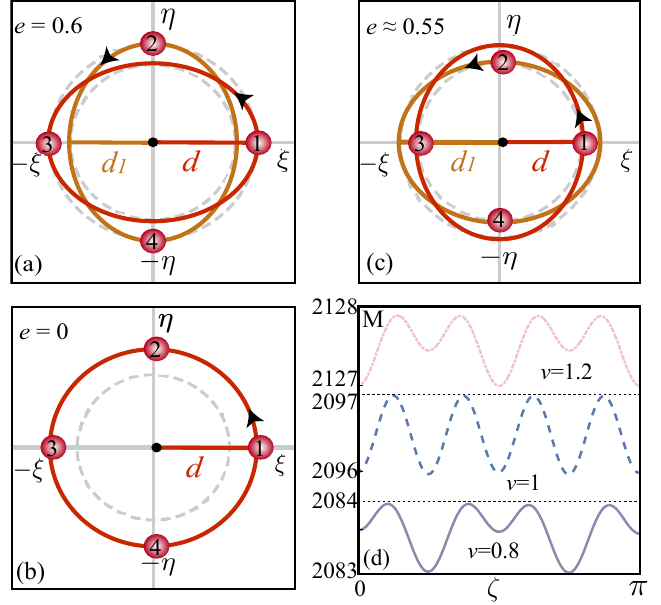}
\caption{{\protect\footnotesize Panel (a), (b),
and (c) display the trajectories corresponding to Figs.~\ref{fig6}(a), (b), and (c),
respectively: (a) and (c) are ellipses with eccentricities 0.6 ($v=0.8$) and 0.55 ($v=1.2$),
while (b) is a circle with $e=0$ ($v=1$). Panel (d) displays the evolution 
of the angular momentum corresponding to (a) $v=0.8$ [purple solid line], (b) $v=1$ [blue dashed line], and $v=1.2$ [red dash-dotted line].
}}
\label{fig7}
\end{figure}
%%%%%%%%%%%%%%%%%%%%%%%%%%%%%%%

In Fig.~\ref{fig6}(a), initial rotation velocity $v=0.8$ is applied to the
SMs. In the course of the propagation, the constituent solitons spiral
inward, their orbital radius continuously decreasing until it attains a
minimum at $\zeta =\pi /4$. In contrast, in the interval $\pi /4<\zeta <\pi
/2$, the solitons spiral outward, while their orbital radius grows until $%
\zeta =\pi /2$, where it returns to the initial value. In the propagation
diagram (the first and third cross sections), we observe that a single
soliton has performed half a period of the rotation during this propagation
interval. The motion then repeats the original sequence from $0$ to $\pi /2$%
, giving the breathing oscillations a period of $\pi /2$. We therefore refer
to these bound states as BMs (\textquotedblleft breather
molecules\textquotedblright ) \cite{xu2019breather,cui2023,Peng2021}.

The breathing period can be obtained, from the equation of motion for the
first soliton in the SM, i.e., $v^{2}\xi _{1}^{2}+\eta _{1}^{2}=v^{2}d^{2}$.
It shows that distance $d$ from the center of mass of the first soliton, $%
(\xi _{1},\eta _{1})$, to the origin of the coordinate system varies as
% $%d\propto \lbrack \cos ^{2}(\beta _{0}\zeta ),\sin ^{2}(\beta _{0}\zeta
% )]^{1/2}$.
$d\propto \left\vert \cos (\beta _{0}\zeta)\right\vert ,\left\vert \sin (\beta _{0}\zeta )\right\vert$.
Here, $\beta _{0}$ is the propagation constant, determined by the system's
parameters, such as atomic density, detunings, and the strength of the
Rydberg interaction. The system's parameters given above yield $\beta _{0}=2$%
. In this case, the breathing oscillation period of the single soliton is $%
T_{b}=\pi /\beta _{0}=\pi /2$. By choosing different values of the system
parameters, one can change the propagation constant $\beta _{0}$ and the
respective breathing period. Thus, our study suggests a scheme for
manipulating the breathing period of the BM.

The last cross-section in Fig. \ref{fig6}(a) shows the trajectory of the
first soliton. We find
that, when $d=\pi $, the first soliton returns to its original position,
indicating that the rotation period is $\pi $, and Fig.~\ref{fig7}(a) presents the
corresponding trajectories of four solitons from Fig.~\ref{fig6}(a). The red
and yellow ellipses with\ eccentricity $e=0.6$ denote, severally, the
trajectories of the first and third solitons, and of the second and fourth
ones. The rotation period can likewise be obtained from the analytical
results. The equations of motion for the center of {mass} of each individual
soliton in the SM in produced in Appendix~\ref{ap1}, see Eqs.~(\ref{eq:sub1}%
) and~(\ref{eq:sub2}). These equations show that the center-of-mass
coordinates $(\xi _{n},\eta _{n})$ of the $n$-th soliton vary sinusoidally
as the function of the propagation distance $\zeta $, i.e., $\xi _{n}(\eta
_{n})\propto \lbrack \cos (\beta _{0}\zeta ),\sin (\beta _{0}\zeta )]$.
Consequently, the SM\ rotation period is $T_{r}=2\pi /\beta _{0}=\pi $. The
propagation constant $\beta _{0}$ can be adjusted by tuning the system's
parameters, thereby enabling control over the rotation period.

When $v=1$, the SM exhibits no radial motion (neither centripetal nor
centrifugal), simply rotating along its initial orbit, and the rotation
period is $\pi $, see Fig.~\ref{fig6}(b). Moreover, the projected SM\
trajectory remains on the circle at all times shown in the last
cross-section in Fig.~\ref{fig6}(b). Figure~\ref{fig7}(b) shows the trajectory of four
solitons corresponding to Fig.~\ref{fig6}(b), which is a circle with
eccentricity $e=0$.

As shown in Fig.~\ref{fig6}(c), with initial rotation velocity $v=1.2$, the
SMs spiral outward, their orbital radius increasing until it reaches a
maximum at $\zeta =\pi /4$. Subsequently, in the interval $\pi /4<\zeta <\pi
/2$, the SMs spiral inward, their orbital radius decreasing until $\zeta
=\pi /2$, where it returns to the initial radius. The motion then repeats
itself, giving the breathing period of $\pi /2$. Figure~\ref{fig7}(c) shows
that the trajectory of the four solitons corresponding to Fig.~\ref{fig6}(c)
is an ellipse with eccentricity $e\approx 0.55$. Thus, adjusting the initial
rotation velocity, one can not only control the trajectories of SMs and BMs,
but also perform mutual conversion between them.

The initial angular velocity gives rise to rotation of the SM, the respective angular momentum $M$ being determined 
by the magnitude of the initial angular velocity. Figure~\ref{fig7}(d) shows the
variation of $M$ in the course of the propagation, for different initial angular velocities. Here, purple solid, blue dashed, and red dash-dotted lines correspond to initial angular velocities $v = 0.8,\, 1$, and 1.2, respectively.

%===========================fig===============================%
\begin{figure}[ht]
\centering
\includegraphics[width=1\linewidth]{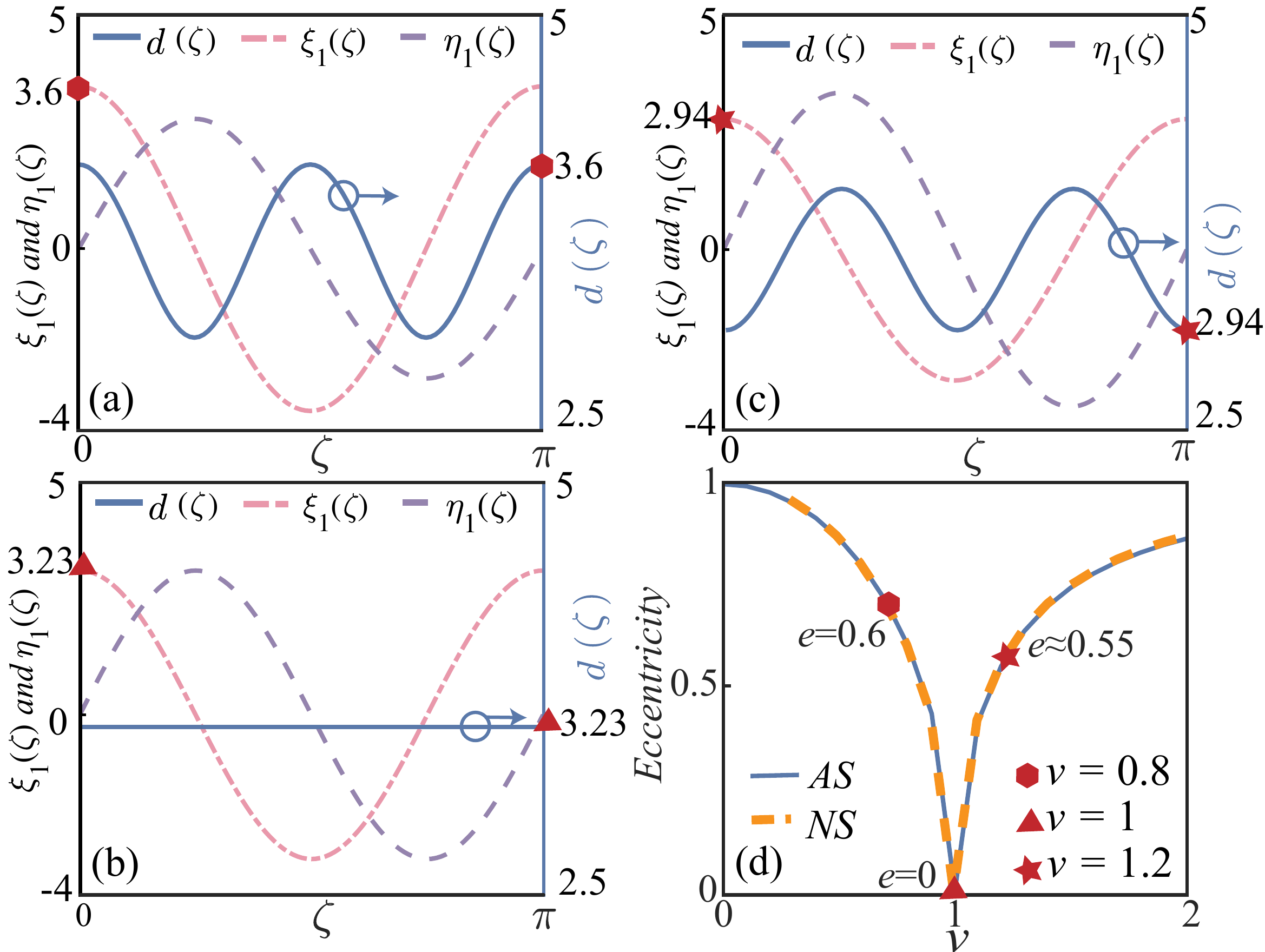}
\caption{{\protect\footnotesize Trajectory projection and eccentricity. (a),
(b), and (c) correspond to the trajectory projections of the single soliton
located at $(\protect\xi _{1},\protect\eta _{1})=(d,0)$ under three
different velocities $v=0.8,\,1,\,1.2$ shown in Fig.~\protect\ref{fig6}. The
solid lines depict the evolution of the center-of-mass distance of first
soliton; the red dotted-dashed and purple dashed lines illustrate the
trajectory projections of the $\protect\xi _{1}-$ and $\protect\eta _{1}-$%
coordinates. The marked points indicate the initial positions and $\protect%
\zeta =\protect\pi $. (d)~Displays the dependence of eccentricity on
velocity $v$. The solid line represents the analytical solution (AS), while
the dotted line corresponds to numerical simulation (NS). In panel (d), the
three marks correspond to the eccentricities of the trajectories in panels
(a), (b), and (c), respectively. }}
\label{fig8}
\end{figure}

To further verify the SM\ and BM\ trajectories, we analyzed equations of
motion for individual solitons. We find that, when $v=0.8$, distance $d$
from the soliton to the origin undergoes periodic variations, first
decreasing and then increasing, see Fig.~\ref{fig8}(a), the $\xi _{1}$ and $%
\eta _{1}$ coordinates of the soliton's center of mass oscillating as cosine
and sine, respectively. These results are consistent with those displayed in
Fig.~\ref{fig6}(a). Concerning Fig.~\ref{fig6}(b), with the initial rotation
velocity $v=1$, we have found that the soliton's distance from the origin
remains fixed on the circle, with its center-of-mass coordinates $\xi _{1}$
and $\eta _{1}$ oscillating as cosine and sine, respectively, indicating
that the soliton simply rotates along its initial orbit, with period $\pi $,
see Fig.~\ref{fig8}(b). This result is consistent with Fig.~\ref{fig6}(b)
and the analytical result presented in Appendix~\ref{ap1}, see Eqs.~(\ref%
{eq:sub1}) and (\ref{eq:sub2}). When $v=1.2$, distance $d$ from the soliton
to the origin undergoes periodic variations, first increasing and then
decreasing. Meanwhile, the $\xi _{1}$ and $\eta _{1}$ coordinates of the
soliton's center-of-mass oscillate as cosine and sine, see Fig.~\ref{fig8}%
(c).

Figure~\ref{fig8}(d) displays the relation between the eccentricity and
rotation velocity. The solid line represents the analytically derived
variation of the eccentricity, while the dashed line represents the
respective numerical results, showing good agreement between both~\cite%
{guoLargePhaseShift2004}. The three markers correspond to Figs.~\ref{fig8}%
(a), (b), and (c), with eccentricities $e=0.6,\,0,$ and $0.55$,
respectively. No numerical data are presented in the range of $0<v<0.3$ due
to the absence of a local BE\ minimum, indicating that there are no stable
SMs or BMs in this velocity range.

\section{Conclusion}

\label{sec5}

Numerous experimental observations of SMs have been
reported in ultrafast fiber lasers operating in the mode-locking regime \cite%
{ortacObservationSolitonMolecules2010, Zavyalov2009, Tang-1, Komarov2009,
qinObservationSolitonMolecules2018, Wang2019, Igbonacho2019,
tsatourian_polarisation_2013, cui2023, Peng2021}.
% Ultrafast fibre lasers, characterized by ultrashort pulse duration and broad spectral bandwidth, which support optical solitons, offer the flexibility for reconfiguring ultrashort pulses.
% By carefully controlling cavity parameters, such as the dispersion, nonlinearity, gain, and loss, these lasers can generate a variety of soliton states~\cite{mao2025}.
% most experimentally observed SMs have been realized in fiber systems. One-dimensional temporal SMs can be spontaneously realized in fiber systems through noise~\cite{Liuxueming2018,Ryczkowski2018}.
Naturally, fiber systems support SMs solely in the 1D form. In this
work, we elaborate the scheme the creation of effectively (2+1)D
SMs in ultracold Rydberg atomic gases. To construct SMs, the first step is
to create multi-soliton clusters. To this end, one can use a
phase-imprinting technique, in which the laser beam is transmitted through
an appropriately oriented set of microscopic glass slides (a phase mask)~%
\cite{Desyatnikov2002}. A four-soliton cluster can be created by means of
two perpendicularly oriented glass slides, with an appropriate tilt inducing
a $\pi $ shift between adjacent solitons. Subsequently, the four-soliton
structure with random noise is transmitted to the input face of the atomic
cell. Under the action of the long-range nonlocal Rydberg interactions,
multiple solitons are locked at a specific equilibrium positions, forming
stable SMs.

In summary, we have proposed a theoretical scheme for realizing SMs and BMs
 in the ultracold Rydberg atomic gas with EIT. In this system, the interplay
between EIT and strong long-range Rydberg interactions generates the
well-defined nonlocal Kerr nonlinearity, which secures the stability of the
(2+1)-dimensional SMs, that is not possible in the case of the local
nonlinearity. In the system with the moderate degree of the nonlocality, we
have predicted various species of stable non-rotating SMs, including
rectangular and oblique rhombuses, checkerboard cells, and hexagons. In the
regime of strong nonlocality, the introduction of rotation is necessary, to
secure to balance of the strong attraction with the centrifugal force.
Varying the rotation velocity, periodic centrifugal-centripetal
\textquotedblleft breathing\textquotedblright\ dynamics of the solitons is
maintained, leading to the creation of BMs. The present results reveal the
crucial roles of the nonlocality and rotation in the formation of SMs and
BMs, providing a reliable means for controlling the size, rotation, and
breathing dynamics of these multi-soliton bound states. The so engineered
robust bound complexes of optical solitons can find applications in the
design of photonic data-processing and transmission technologies.

As an extension of the analysis, it may be interesting to construct bound
states of several optical solitons with \textit{intrinsic vorticity}. In
particular, it may be possible to set such self-supported complexes of
vortex solitons in overall rotation, thus realizing the concept of
\textquotedblleft supervortices", cf. Ref. \cite{sakaguchi2005higher}.

% \section{EXPERIMENTAL OUTLOOK}
% \label{sec6}
% {\color{blue}In experiments, it is possible to achieve the various configuration SMs described in this paper with the development of modern experimental techniques. We can use EIT to couple the probe light excitation field transition to the Rydberg state, obtaining \(^{88}\mathrm{Sr}\) atoms with a principal quantum number of $n=60$. The specific operation can be referred to in reference~\cite{Pritchard2010}. The experiment uses single-photon level detection to directly observe and precisely characterize this long-range light-light interaction, providing the necessary nonlinear environment for the formation of soliton molecules~\cite{Gorniaczyk2014}. In terms of light field preparation, the high-order Gaussian light array can be obtained using a spatial light modulator (SLM), and this technique has been experimentally achievable~\cite{RubinszteinDunlop2016,Weiner2011}. After coupling the preformed probe light into the above Rydberg-EIT medium, the SMs interact through nonlocal nonlinear sound waves over long distances and spontaneously evolve into stable bound states. In conclusion, by combining the verified Rydberg nonlocal nonlinearities and mature wavefront shaping techniques, the rich and stable high-dimensional SM structures predicted by this paper have a clear realization prospect within the existing experimental technology framework, laying the foundation for future exploration of related applications in the field of optical information processing.}

\acknowledgments
This work was supported by the National Natural Science Foundation of China
(NSFC) under Grants No. 12404377, 12334012, and 62505079; the Key Scientific Research Project of Colleges and Universities in Henan Province (26B140007); and the Key International Cooperation Project in Henan Province (Grant No. 261111521200).

\appendix

\section{The Bloch equations}\label{AP:BE}

The full system of the optical Bloch equations for the evolution of the
density-matrix elements $\rho _{\alpha \beta }\equiv \langle {\hat{S}}%
_{\beta \alpha }\rangle $ is
\begin{subequations}
\label{Bloch}
\begin{align}
& i\frac{\partial }{\partial t}\rho _{11}-i\Gamma _{12}\rho _{22}+\Omega
_{p}^{\ast }\rho _{21}-\Omega _{p}\rho _{12}=0,  \label{11} \\
& i\frac{\partial }{\partial t}\rho _{22}+i\Gamma _{12}\rho _{22}-i\Gamma
_{23}\rho _{33}+\Omega _{c}^{\ast }\rho _{32}-\Omega _{c}\rho _{23}\notag\\
&\quad-\Omega
_{p}^{\ast }\rho _{21}+\Omega _{p}\rho _{12}=0,  \label{Bloch22} \\
& i\frac{\partial }{\partial t}\rho _{33}+i\Gamma _{23}\rho _{33}-\Omega
_{c}^{\ast }\rho _{32}+\Omega _{c}\rho _{23}=0,  \label{Bloch33} \\
& \left( i\frac{\partial }{\partial t}+d_{21}\right) \rho _{21}+\Omega
_{c}^{\ast }\rho _{31}-\Omega _{p}(\rho _{22}-\rho _{11})=0,  \label{Bloch21}
\\
& \left( i\frac{\partial }{\partial t}+d_{31}\right) \rho _{31}-\Omega
_{p}\rho _{32}+\Omega _{c}\rho _{21}\notag\\
&\quad-\frac{\mathcal{N}%
_{a}}{2}\int d^{3}\mathbf{%
r^{\prime }}V(\mathbf{r^{\prime }}-\mathbf{r})\rho \rho _{33,31}(\mathbf{%
r^{\prime }},\mathbf{r},t)=0,  \label{31} \\
& \left( i\frac{\partial }{\partial t}+d_{32}\right) \rho _{32}-\Omega
_{p}^{\ast }\rho _{31}-\Omega _{c}(\rho _{33}-\rho _{22})\notag\\
&\quad-\frac{\mathcal{N}%
_{a}}{2}\int d^{3}\mathbf{r^{\prime }}V(\mathbf{r^{\prime }}-\mathbf{r})\rho
\rho _{33,32}(\mathbf{r^{\prime }},\mathbf{r},t)=0,  \label{32}
\end{align}%
\end{subequations}
where $\rho_{\alpha\beta}=\langle\hat{S}_{\beta\alpha}\rangle$ are
one-body density matrix elements,
$d_{\alpha\beta}=\Delta_{\beta}-\Delta_{\alpha}+i\gamma_{\alpha\beta}$~($\Delta_1=0$),
% $d_{21}=\Delta_{2}+i\gamma_{21}$, $%
% d_{31}=\Delta _{3}+i\gamma _{31}$, $d_{32}=\Delta_{3}-\Delta_{2}+i\gamma
% _{32}$,
$\gamma_{\alpha \beta }=(\Gamma _{\alpha }+\Gamma _{\beta
})/2+\gamma_{\alpha \beta }^{\mathrm{dep}}$ ($\alpha \neq \beta $), and $%
\Gamma_{\beta}=\sum_{\alpha <\beta }\Gamma _{\alpha \beta }$, with $\Gamma
_{\alpha \beta}$ and $\gamma_{\alpha \beta }^{\mathrm{dep}}$ being the
spontaneous decay and dephasing rates, respectively, for the $|\beta \rangle
\rightarrow |\alpha \rangle $ transition.

In the last term on left-hand side of Eqs.~(\ref{31}) and (\ref{32}), the
two-body correlators $\rho _{33,3\alpha }(\mathbf{r^{\prime },r},t)\equiv
\langle {\hat{S}}_{33}(\mathbf{r^{\prime }},t){\hat{S}}_{3\alpha }(\mathbf{r}%
,t)\rangle $ $(\alpha =1,\,2)$ originate from the Rydberg-Rydberg interaction.

\section{The solution of the Bloch equations}\label{AP:WG}

We assume that the probe field is much weaker than the control field and
atoms initially populate state $|1\rangle $. Thus, one can use the
asymptotic expansion for the density matrix: $\rho _{\alpha \beta }=\epsilon
\rho _{\alpha \beta }^{(1)}+\epsilon ^{2}\rho _{\alpha \beta
}^{(2)}+\epsilon ^{3}\rho _{\alpha \beta }^{(3)}\cdots $ ($\alpha,\,\beta
=1,\,2,\,3$), $\Omega _{p}=\sum_{l=1}\epsilon ^{l}\Omega _{p}^{(l)}$, where $%
\epsilon $ a parameter characterizing the amplitude of $\Omega _{p}$. The
terms of the expansion are considered as functions of multiscale variables, $%
z_{l}=\epsilon ^{l}z$ $(l=0,\,1,\,2)$,  and $(x_{1},y_{1})=\epsilon (x,y)$. Substituting the expansion in the
MB equations, and comparing coefficients in front of $\epsilon
^{l}~(l=1,\,2,\cdots )$, we obtain a set of linear inhomogeneous equations
which can be solved order by order.
Notice that in this work we are interested in the
steady-state properties of the system, for which the time derivative in the
MB equations may be neglected, which is valid for the probe field with a
long temporal duration.

At the first order ($l=1$), we obtain the solution
\begin{subequations}
\begin{align}
& \rho _{21}^{(1)}=\frac{d_{31}}{D_1},~\rho _{31}^{(1)}=-\frac{\Omega _{c}}{D_1},\notag
\end{align}%
\end{subequations}
where $D_1=|\Omega _{c}|^{2}-d_{21}d_{31}$.

For the second order ($l = 2$), one obtains the solution
\begin{subequations}
\begin{align}
\rho_{11}^{(2)} &= \frac{[i\Gamma_{23} - 2|\Omega_c|^2 M_1]M_2 - i\Gamma_{12}|\Omega_c|^2M_3} {\Gamma_{12}\Gamma_{23} - i\Gamma_{12}|\Omega_c|^2 M_1}, \\
%%%
\rho_{33}^{(2)} &= \frac{1}{i\Gamma_{12}}\left(M_2 - i\Gamma_{12}\rho_{11}^{(2)}\right), \\
%%%%
\rho_{32}^{(2)} &= \frac{1}{d_{32}}\left(-\frac{\Omega_c}{D_1} + 2\Omega_c\rho_{33}^{(2)} + \Omega_c\rho_{11}^{(2)}\right),
\end{align}
\end{subequations}
where $M_1 = 1/d_{32} - 1/d_{32}^*$, $M_2 = d_{31}^*/D_1^* - d_{31}/D_1$, and $M_3=1/(D_1^*d_{32}^*)-1/(D_1 d_{32})$.

At the third order ($l = 3$), the solution reads
\begin{subequations}
\begin{align}
&\rho_{21}^{(3)} = {\mathcal A}_{21}^{(3)} + \mathcal{N}_a {\mathcal B}_{21}^{(3)},\\
&\rho_{31}^{(3)} = {\mathcal A}_{31}^{(3)} + \mathcal{N}_a {\mathcal B}_{31}^{(3)},
\end{align}
\end{subequations}
with
%%%%
\begin{subequations}
\begin{align}
{\mathcal A}_{21}^{(3)} &= \frac{\Omega_c^*\rho_{32}^{(2)} + d_{31}(2\rho_{11}^{(2)} + \rho_{33}^{(2)})}{D_1},\\
{\mathcal B}_{21}^{(3)} &= \frac{\Omega_c^*\int d^3\mathbf{r}'  \rho_{33,31}^{(3)}(\mathbf{r}' - \mathbf{r})V(\mathbf{r}' - \mathbf{r})}{D_1}, \\
%%%%%%%%%
{\mathcal A}_{31}^{(3)} &= \frac{-(2\rho_{11}^{(2)} + \rho_{33}^{(2)})\Omega_c + d_{21}\rho_{32}^{(2)}}{D_1},\\
{\mathcal B}_{31}^{(3)} &= \frac{\int d^3\mathbf{r}'  \rho_{33,31}^{(3)}(\mathbf{r}' - \mathbf{r})V(\mathbf{r}' - \mathbf{r})d_{21}}{D_1},
\end{align}
\end{subequations}
where $\rho _{33,31}^{(3)}(\mathbf{r^{\prime }-r})=\langle {\hat{S}}_{33}(%
\mathbf{r^{\prime }}){\hat{S}}_{31}(\mathbf{r})\rangle $ is the two-body
correlator.

According to the third-order solution,
expressions for coefficients of the local and nonlocal optical Kerr
nonlinearity are
\begin{subequations}
\begin{align}
& W=\kappa _{12}\mathcal{A}_{21}^{(3)}, \\
& G=\kappa _{12}\mathcal{N}_{a}\mathcal{B}_{21}^{(3)}.\label{GG}
\end{align}
\end{subequations}
% \begin{subequations}
% \begin{align}
% & W=\kappa _{12}\frac{d_{31}(2a_{11}^{(2)}+a_{33}^{(2)})+\Omega
% _{c}^{\ast }a_{32}^{(2)}}{D}, \\
% & G=\kappa _{12}\frac{\Omega _{c}^{\ast }\mathcal{N}_{a}a_{33,31}^{(3)}(%
% \mathbf{r^{\prime }-r})V(\mathbf{r^{\prime }-r})}{D}.
% \end{align}
% \end{subequations}
Because the dephasing in the system is much weaker than the spontaneous emission, Eq.~(\ref{GG}) can be approximately written as~\cite{bai2019stable,qin2022stable}
\begin{align}
G\approx  \kappa_{12}\mathcal{N}_a\frac{2(d_{21} + d_{31})|\Omega_c|^4/|D_1|^2}{ D_1D_2[2d_{31} - V(\mathbf{r}' - \mathbf{r})]-2d_{21}|\Omega_c|^2},\notag
\end{align}
% \begin{align}
%  a_{33,31}^{(3)}\approx  \frac{2(d_{21} + d_{31})|\Omega_c|^2\Omega_c/|D|^2}{ E[2d_{31} - V(\mathbf{r}' - \mathbf{r})]-2d_{21}|\Omega_c|^2},\notag
% \end{align}
with $D_2 = |\Omega_c|^2 - d_{21}(d_{21} + d_{31})$.

\section{The NLS equation in the regime of strongly nonlocal nonlinearity}

\label{AP:A}

Assuming that the pulse's temporal duration is sufficiently long to make the
group-velocity dispersion of the system negligible, the light propagation in
the spatial domain is governed by Eq. (\ref{nls}). In the case of strong
nonlocality, the response function $G$ in Eq. (\ref{nls}) is very flat near $%
(x^{\prime },y^{\prime })=(x,y)$, in comparison to the intensity profile of
the probe field, which allows one to use the Taylor's expansion,  $%
G(x-x^{\prime },y-y^{\prime })\approx G(0)+G^{\prime \prime
}(0)(x^{2}+y^{2})/2$ \cite{bai2019stable,snyderAccessibleSolitons1997}. In
this case, Eq.~(\ref{nls}) can be written as
\begin{align}
i\frac{\partial }{\partial z}\Omega _{\mathrm{p}}+& \frac{c}{2\omega _{%
\mathrm{p}}}\nabla _{\perp }^{2}\Omega _{\mathrm{p}}+W|\Omega _{\mathrm{p}%
}|^{2}\Omega _{\mathrm{p}}  \notag \\
+& P_{0}[G(0)+\frac{1}{2}G^{\prime \prime }(0)(x^{2}+y^{2})]\Omega _{\mathrm{p}}=0,
\label{eq10}
\end{align}%
where $P_{0}=\iint dx\,dy\,|\Omega _{\mathrm{p}}|^{2}$ is the power of the
probe field.
Because the expression for the nonlocal response function $G(x-x',y-y')$ is very complex, 
the coefficients \(G(0)\) and \(G_2^{\prime \prime }(0)\) were obtained in the numerical form.
Equation~(\ref{eq10}) can be cast in the scaled form,
\begin{equation}
i\frac{\partial u}{\partial \zeta }+\left( \frac{\partial ^{2}}{\partial \xi
^{2}}+\frac{\partial ^{2}}{\partial \eta ^{2}}\right)
u+w|u|^{2}u+g_{1}u+g_{2}(\xi ^{2}+\eta ^{2})u=0,  \label{gg}
\end{equation}%
where $w=2W|u_{0}|^{2}L_{\mathrm{diff}}$, $g_{1}=2L_{\mathrm{diff}}P_{0}G(0)$%
, $g_{2}=L_{\mathrm{diff}}R_{0}^{2}P_{0}G^{\prime \prime }(0)$. Thus, in the
case of the strongly nonlocal nonlinear response region, the nonlocal part
of the governing NLS\ equation (\ref{gg}) amounts to the quasi-linear form
with the harmonic-oscillator potential, similar to the model of
\textquotedblleft accessible soliton" \cite{snyderAccessibleSolitons1997}.
By substituting $\Phi =u\exp (ig_{1}\zeta )$, Eq. (\ref{gg}) is further
reduced to Eq. (\ref{Phi}) in the main text, which makes it possible to look
for solutions in the form of Gaussians (approximate ones, if Eq. (\ref{gg})
keeps $w\neq 0$) \cite{snyderAccessibleSolitons1997,song2018controllable}.

\section{The analytical solution for SMs}

\label{ap1}

In the strongly nonlocal regime, Eq.~(\ref{Phi}) with $w=0$ has an exact
Gaussian solution~\cite{guoLargePhaseShift2004}, which is actually
tantamount to the well known exact solution of the Schr\"{o}dinger
equation, representing a wave function of the harmonic oscillator in quantum mechanics \cite%
{bohm2012quantum}
\begin{equation}
\Phi (\mathbf{r},\zeta )=A\exp \left[ -\frac{\xi ^{2}+\eta ^{2}}{%
2a^{2}(\zeta )}+ib(\zeta )\left( \xi ^{2}+\eta ^{2}\right) +i\theta (\zeta )%
\right] ,  \notag
\end{equation}%
where
\begin{align}
a(\zeta )& =a_{0}\left( \cos ^{2}\alpha +P_{r}\sin ^{2}\alpha \right) ^{1/2},
\notag \\
b(\zeta )& =\frac{\beta _{0}(P_{r}-1)\sin (2\alpha )}{4\left( \cos
^{2}\alpha +P_{r}\sin ^{2}\alpha \right) },  \notag \\
\theta (\zeta )& =-\arctan \left( \sqrt{P_{r}}\tan \alpha \right) ,  \notag
\end{align}%
denote the beam's width, phase-front curvature (chirp) \cite{desaix1991variational}, and
overall phase, respectively. Further, $a_{0}=a(0)$ is the initial width of
the beam, and $\alpha =\beta _{0}\zeta $, where $\beta _{0}=2\sqrt{|L_{%
\mathrm{diff}}R_{0}^{2}P_{0}G^{\prime \prime }(0)|}$ is the propagation
constant. We define $P_{r}=P_{c}/P_{0}$ as the power ratio, $P_{c}=1/|L_{%
\mathrm{diff}}R_{0}^{2}G^{\prime \prime }(0)a^{4}|$ being the critical
power. When $P_{r}=1$ $(P_{0}=P_{c})$, the chirp-induced compression exactly
balances the diffraction, hence the respective Gaussian beam preserves its
width in the course of the propagation, being tantamount to the
ground-state wave function of the 2D harmonic oscillator in quantum
mechanics. With $P_{r}\neq 1$, the width of the Gaussian beam periodically
oscillates, making it possible to identify it as quasi-linear breather \cite%
{snyderAccessibleSolitons1997,song2018controllable}. We primarily focus on
the case $P_{0}=P_{c}$, when the beam keep its steady state.

If $\Phi (\mathbf{r},\zeta )$ is a solution of Eq.~(\ref{Phi}), then
\begin{equation}
\Phi _{n}(\mathbf{r},\zeta )=\Phi (\mathbf{r}-\mathbf{r}_{n}(\zeta ),\zeta
)\exp \left[ i\mathbf{v}_{r}(\zeta )(\mathbf{r}-\mathbf{r}_{n}(\zeta
))+i\Theta (\zeta )\right]\notag
\end{equation}%
are solutions as well, where $\mathbf{r}_{n}(\zeta )$, $\mathbf{v}_{r}(\zeta
)$, $\Theta (\zeta )$ satisfy the following equations:
\begin{align}
\mathbf{r}_{n}^{\prime \prime }(\zeta )& +\beta _{0}^{2}\mathbf{r}_{n}(\zeta
)=0,  \notag \\
\mathbf{v}_{r}(\zeta )& =\frac{1}{2}\boldsymbol{\mathbf{r}}_{n}^{\prime
}(\zeta ),  \notag \\
\Theta ^{\prime }(\zeta )& =\frac{1}{4}\left[ \beta _{0}^{2}\mathbf{r}%
_{n}^{2}(\zeta )-\mathbf{r}_{n}^{\prime 2}(\zeta )\right] .  \notag
\end{align}%
For the quasi-linear equation (\ref{gg}), we construct a coherent
superposition of $N$ Gaussian beams, the respective quasi-linear limit form
of the SM being
\begin{equation}
\Psi _{N}(\mathbf{r},\zeta )=\sum_{n=1}^{N}\Phi _{n}(\mathbf{r},\zeta ).
\label{u}
\end{equation}%
Explicit expressions for $\Phi _{n}(\mathbf{r},\zeta )$ can be written in
% \begin{widetext}
{\small \
\begin{align}
\Phi & _{n}(\mathbf{r},\zeta )=A\exp \biggl\{-\frac{(\xi -\xi
_{n})^{2}+(\eta -\eta _{n})^{2}}{2a^{2}(\zeta )}+ib(\zeta )\left[ (\xi -\xi
_{n})^{2}\right.   \notag \\
& \left. +(\eta -\eta _{n})^{2}\right] +i[\left( v_{\xi }\right) _{n}(\xi
-\xi _{n})+\left( v_{\eta }\right) _{n}(\eta -\eta _{n})]+i\Theta _{n}(\zeta
)+i\theta _{n}\biggr\},  \notag
\end{align}%
} % \end{widetext}
where
\begin{subequations}
\label{eq:group}
\begin{align}
&\xi _{n}(\zeta ) =\xi _{n}(0)\cos \alpha +\frac{\left( v_{\xi }\right)
_{n}(0)}{\beta _{0}}\sin \alpha ,  \label{eq:sub1} \\
&\eta _{n}(\zeta ) =\eta _{n}(0)\cos \alpha +\frac{\left( v_{\eta }\right)
_{n}(0)}{\beta _{0}}\sin \alpha ,  \label{eq:sub2} \\
&\left( v_{\xi }\right) _{n}(\zeta ) =-\frac{1}{2}\xi _{n}(0)\beta _{0}\sin
\alpha +\frac{1}{2}\left( v_{\xi }\right) _{n}(0)\cos \alpha ,
\label{eq:sub3} \\
&\left( v_{\eta }\right) _{n}(\zeta ) =-\frac{1}{2}\eta _{n}(0)\beta
_{0}\sin \alpha +\frac{1}{2}\left( v_{\eta }\right) _{n}(0)\cos \alpha ,
\label{eq:sub4} \\
&\Theta _{n}(\zeta ) =\frac{1}{8}\left[ \beta _{0}\left( \xi
_{n}^{2}(0)+\eta _{n}^{2}(0)\right) -\frac{\left( v_{\xi }\right)
_{n}^{2}(0)+\left( v_{\eta }\right) _{n}^{2}(0)}{\beta _{0}}\right]   \notag \\
& \times\sin
(2\alpha )-\frac{1}{4}\left[ \xi _{n}(0)\left( v_{\xi }\right) _{n}(0)+\eta
_{n}(0)\left( v_{\eta }\right) _{n}(0)\right] \cos (2\alpha ).
\label{eq:sub5}
\end{align}%
\end{subequations}
The initial rotation velocity of the SM\ is defined as $[\left( v_{\xi
}\right) _{n}(0),\,\left( v_{\eta }\right) _{n}(0)]$ = $v\beta _{0}[-\eta
_{n}(0),\,\xi _{n}(0)]$. The initial rotation velocity of each Gaussian
constituent is determined by the condition of obliquely symmetric incidence,
i.e., by angle $\vartheta $ between the wave vector and the propagation
direction at the incidence position~\cite{song2018controllable}. The wave
vector of each constituent at $\zeta =0$ can be written as
\begin{align}
\mathbf{k}_{n}&=k_{n\xi }\mathbf{e}_{\xi }+k_{n\eta }\mathbf{e}_{\eta
}+k_{n\zeta }\mathbf{e}_{\zeta }\notag \\
&=k\left( v_{\xi }\right) _{n}\mathbf{e}_{\xi }+k\left( v_{\eta }\right) _{n}%
\mathbf{e}_{\eta }+k\mathbf{e}_{\zeta },
\end{align}%
where $\mathbf{e}_{j}$ $(j=\xi,\,\eta,\,\zeta )$ is the unit vector in the $j$%
-direction, $k=2\pi n_{0}/\lambda $, and $k_{nj}$ are components of the wave
vector. Then, the angle of the oblique incidence is determined by the
condition
\begin{equation}
\tan (\vartheta _{n})=\frac{k_{\perp }}{k}={\ \sqrt{\left( v_{\xi }\right)
_{n}^{2}+\left( v_{\eta }\right) _{n}^{2}}},~\quad k_{\perp }=\sqrt{k_{n\xi
}^{2}+k_{n\eta }^{2}},
\end{equation}%
hence $\vartheta _{n}=\arctan \left( \sqrt{\left( v_{\xi }\right)
_{n}^{2}+\left( v_{\eta }\right) _{n}^{2}}\right) $.

%===========================fig===============================%
\begin{figure}[ht]
\centering
\includegraphics[width=1\linewidth]{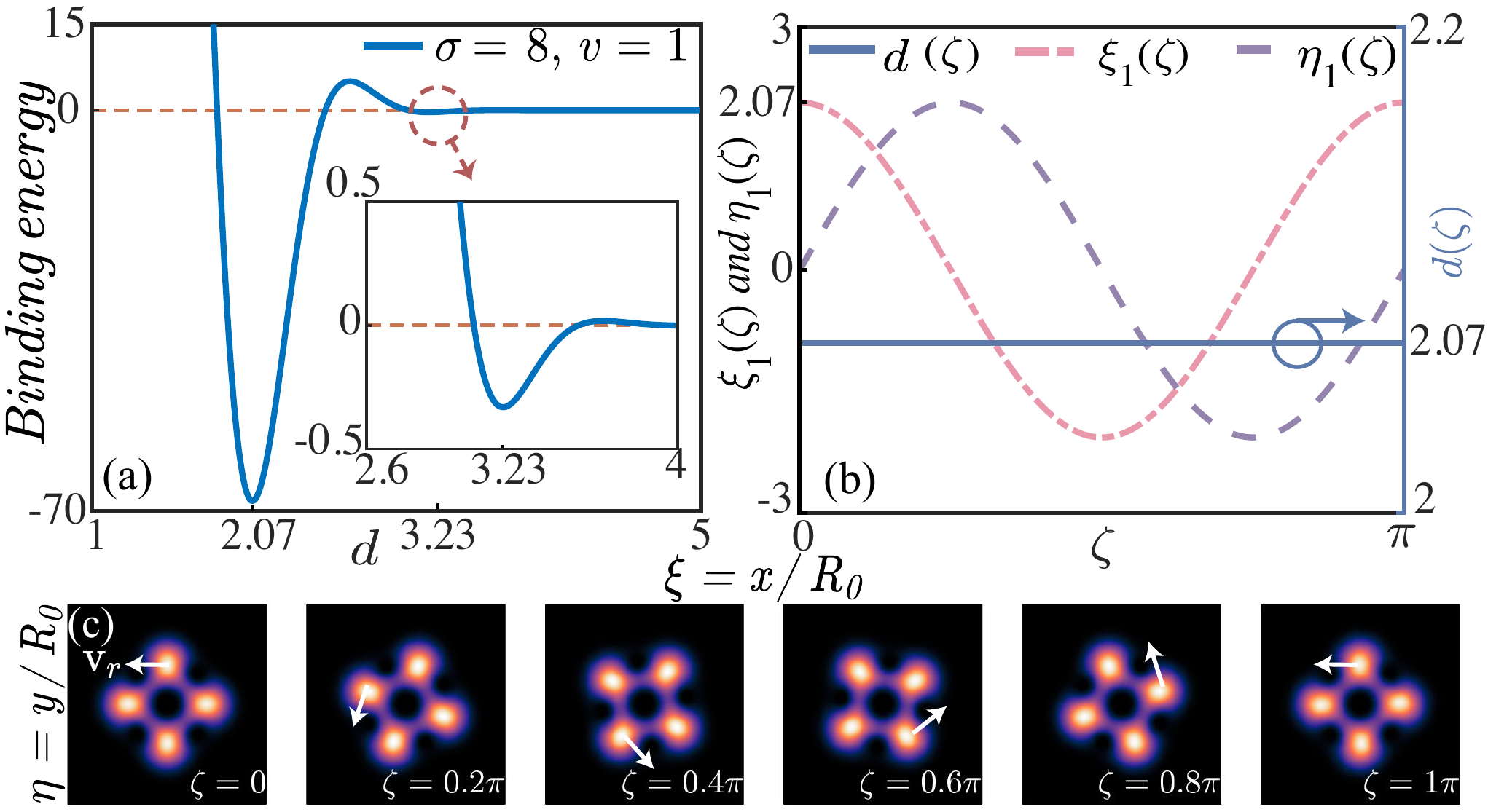}
\caption{{\protect\footnotesize (a)~The BE exhibits two local minima at $%
d=2.07$ and $d=3.23$ in the strongly nonlocal regime. Inset: The second
minimum in the BE aligns with the purple dashed-dotted line shown in Fig.~%
\protect\ref{fig2}(c). (b)~The trajectory of the single soliton located at $(%
\protect\xi _{1},\protect\eta _{1})=(d,0)$ with $d=2.07$ is presented,
along with the evolution of the center-of-mass position $(\protect\xi _{1},%
\protect\eta _{1})$ and $d$, as functions of the propagation distance.
(c)~With the initial rotation velocity $v=1$, the SM exhibits a
counterclockwise rotational motion. The white arrows indicate how the
tangent velocity of a particular soliton evolves during the propagation. As
seen in the figure, adjacent solitons are connected by mutual interactions,
the rotation period of the soliton being $\protect\pi $. The other
parameters are the same as Fig.~\protect\ref{fig2}(c). }}
\label{fig9}
\end{figure}

\section{Contact-type SMs}

\label{AP:C}

According to the energy functional and BE equations, we find that more than
one equilibrium point exists in the strongly nonlocal nonlinear regime. Here
we focus on the first equilibrium position. When the soliton is placed at
this position, with $d=2.07$, and set in rotation by the initial tangent
velocity $v=1$, the resulting structure is a contact-type SM, as shown in
Fig.~\ref{fig9}(c). The other parameters $A=4$, $a=1$, and $b=0$.

Next, we consider the SM's trajectory. Combining Eqs. (\ref{eq:sub1}) and (%
\ref{eq:sub2}), the trajectory of the center-of-mass of each constituent
soliton in the $\left( \xi ,\eta \right) $ plane can be obtained:
\begin{align}
& \left[ \left( v_{\eta }\right) _{n}(0)\xi _{n}-\left( v_{\xi }\right)
_{n}(0)\eta _{n}\right] ^{2}+\beta _{0}^{2}\left[ \eta _{n}(0)\xi _{n}-\xi
_{n}(0)\eta _{n}\right] ^{2}  \notag \\
&\quad =\left[ \xi _{n}(0)\left( v_{\eta }\right) _{n}(0)-\eta
_{n}(0)\left( v_{\xi }\right) _{n}(0)\right] ^{2}.  \label{Eq:trajectory}
\end{align}%
Taking the first soliton, initially located at $(\xi _{1},\eta _{1})=(d,0)$,
and substituting the input conditions in Eq.~(\ref{Eq:trajectory}), we
obtain the standard ellipse:
\begin{equation}
v^{2}\xi _{1}^{2}(\zeta )+\eta _{1}^{2}(\zeta )=d^{2}v^{2}.
\end{equation}%
Actually, all constituent solitons move along the same trajectory.

When $v=1$, Eq.~\eqref{Eq:trajectory} becomes the circle equation, $\xi
_{1}^{2}+\eta _{1}^{2}=d^{2}$, with the circle's radius exactly equal to
that of the initial ring. As shown by the blue solid line in Fig.~\ref{fig9}%
(b), the circle's radius remains constant. The red doted-dashed and purple
dashed lines trace the $\xi _{1}$ and $\eta _{1}$ trajectories, executing
one full cosine and sine cycle, respectively, to confirm that the soliton
follows a closed circular orbit.

In Fig.~\ref{fig9}(c), one sees that adjacent solitons in the SM are
connected through the mutual interactions, i.e., these are contact-type SM.
As the tangent velocity is directed counterclockwise, the soliton SM rotates in
that direction. The white arrows indicate how the soliton's tangent velocity
evolves in the course of the propagation. As observed in the figure, the
rotation period of each soliton is $\pi $.

% ---- 参考文献 ----
% \bibliographystyle{plain}
\bibliography{allin}
% 需要references.bib文件

% {\LARGE Please add these items to the bibliography:}

% \begin{thebibliography}{9}
% \bibitem{book} B. A. Malomed, \textit{Multidimensional Solitons} (AIP\
% Publishing, Melville, NY, 2022).

% \bibitem{Avi} I. Tikhonenkov, B. A. Malomed, and A. Vardi, Anisotropic
% solitons in dipolar Bose-Einstein condensates, Phys. Rev. Lett. \textbf{100}%
% , 090406 (2008).

% \bibitem{Desaix} M. Desaix, D. Anderson, and M. Lisak, Variational approach
% to collapse of optical pulses, J. Opt. Soc. Am. B \textbf{8}, 2082-2086
% (1991).

% \bibitem{bohm2012quantum} D. Bohm, \textit{Quantum Theory}, Dover, New York, 1989.

% \bibitem{HS} H. Sakaguchi and B. A. Malomed, Higher-order vortex solitons,
% multipoles, and supervortices on a square optical lattice, Europhys. Lett.
% \textbf{72}, 698-704 (2005).
% \end{thebibliography}

\end{document}